\documentclass[aps,pre,11pt,notitlepage,tightenlines,eqsecnum,longbibliography]{revtex4-1}

\usepackage{amsmath}
\usepackage{amssymb,amsfonts}
\usepackage{wasysym}
\usepackage{graphicx}
\usepackage{tikz}
\usepackage{bm}
\usepackage{subfigure}
\usepackage{color}
\usepackage{hyperref}

\graphicspath{{figs/}}

\usepackage[letterpaper]{geometry}

\newcommand{\mathnotation}[2]{\newcommand{#1}{\ensuremath{#2}}}
\newcommand{\nofrac}[2]{#1/#2}

\renewcommand{\l}{\left}
\renewcommand{\r}{\right}
\mathnotation{\pd}{\partial}
\mathnotation{\ldef}{\mathrel{\raisebox{.069ex}{:}\!\!=}}
\mathnotation{\rdef}{\mathrel{=\!\!\raisebox{.069ex}{:}}}
\mathnotation{\dint}{\,{\mathrm{d}}}
\mathnotation{\ee}{\mathrm{e}}      

\renewcommand{\t}{t}                
\mathnotation{\xc}{x}               
\mathnotation{\yc}{y}               
\mathnotation{\xv}{\bm{\xc}}        
\mathnotation{\dvisc}{\mu}          
\mathnotation{\uc}{u}               
\mathnotation{\uv}{\bm{\uc}}        
\mathnotation{\U}{U}                
\mathnotation{\eye}{\mathbb{I}}     
\mathnotation{\s}{s}                
\mathnotation{\shat}{\hat{\bm{\s}}} 
\mathnotation{\fc}{f}               
\mathnotation{\fv}{\bm{\fc}}        
\mathnotation{\T}{T}                
\mathnotation{\TL}{\T_{\L}}         
\mathnotation{\Tz}{\T_0}            
\mathnotation{\E}{E}                
\mathnotation{\I}{I}                
\renewcommand{\L}{L}                
\mathnotation{\X}{X}                
\mathnotation{\Xv}{\bm{\X}}         
\mathnotation{\eps}{\varepsilon}    
\mathnotation{\R}{R}                
\mathnotation{\rr}{r}               
\mathnotation{\dd}{d}               
\mathnotation{\Lmax}%
        {\L_{\text{max}}}           
\mathnotation{\DLmax}{\Delta\Lmax}  
\mathnotation{\tdep}%
        {\t_{\text{dep}}}           
\mathnotation{\tdepfifty}%
        {\t_{\mathrm{dep},50\%}}    
\mathnotation{\V}{V}                
\mathnotation{\FVconst}{\alpha}     
\mathnotation{\FVpow}{m}            
\mathnotation{\peel}{{\mathrm{P}}}  
\mathnotation{\Fp}{F_\peel}         
\mathnotation{\Ap}{A_\peel}         
\mathnotation{\Lp}{\L_\peel}        
\mathnotation{\dLp}{\dot{\L}_\peel} 
\mathnotation{\Wp}{W_\peel}         
\mathnotation{\deltap}{\delta_\peel}
\mathnotation{\drag}{{\mathrm{D}}}  
\mathnotation{\Fd}{F_\drag}         
\mathnotation{\dotEner}%
        {{\dot E}_{\text{total}}}   
\mathnotation{\wpmin}%
        {\wp_{\text{min}}}          
\mathnotation{\Lc}{L_c}             

\newcommand{\microm}{\ensuremath{\mu\mathrm{m}}}
\newcommand{\cm}{\ensuremath{\mathrm{cm}}}
\newcommand{\meter}{\ensuremath{\mathrm{m}}}
\newcommand{\Pa}{\ensuremath{\mathrm{Pa}}}

\newcommand{\milliPa}{\ensuremath{\mathrm{mPa}}}
\newcommand{\second}{\ensuremath{\mathrm{s}}}
\newcommand{\millisecond}{\ensuremath{\mathrm{ms}}}
\newcommand{\Newton}{\ensuremath{\mathrm{N}}}

\newcommand{\microNewton}{\ensuremath{\mu\mathrm{N}}}

\newcommand{\mJoule}{\ensuremath{\mathrm{mJ}}}

\begin{document}

\title{Unraveling hagfish slime}

\author{Gaurav Chaudhary}
\email{gchaudh2@illinois.edu}
\affiliation{Department of Mechanical Science and Engineering, University of Illinois at Urbana-Champaign, 1206 W. Green St., Urbana, IL 68101, USA}

\author{Randy H. Ewoldt}
\email{ewoldt@illinois.edu}
\affiliation{Department of Mechanical Science and Engineering, University of Illinois at Urbana-Champaign, 1206 W. Green St., Urbana, IL 68101, USA}

\author{Jean-Luc Thiffeault}
\email{jeanluc@math.wisc.edu}
\affiliation{Department of Mathematics, University of Wisconsin --
  Madison, 480 Lincoln Dr., Madison, WI 53706, USA}

\date{\today}

\begin{abstract}
Hagfish slime is a unique predator defense material containing a network of long fibrous threads each $\sim 10\,\cm$ in length. Hagfish release the threads in a condensed coiled state known as thread cells, or skeins ($\sim 100\,\microm$), which must unravel within a fraction of a second to thwart a predator attack. Here we consider the hypothesis that viscous hydrodynamics can be responsible for this rapid unraveling, as opposed to chemical reaction kinetics alone. Our main conclusion is that, under reasonable physiological conditions, unraveling due to viscous drag can occur within a few hundred milliseconds, and is accelerated if the skein is pinned at a surface such as the mouth of a predator.  We model a single thread cell unspooling as the fiber peels away due to viscous drag. We capture essential features by considering one-dimensional scenarios where the fiber is aligned with streamlines in either uniform flow or uniaxial extensional flow. The peeling resistance is modeled with a power-law dependence on peeling velocity. A  dimensionless ratio of viscous drag to peeling resistance appears in the dynamical equations and determines the unraveling timescale. Our modeling approach is general and can be refined with future experimental measurements of peel strength for skein unraveling. It provides key insights into the unraveling process, offers potential answers to lingering questions about slime formation from threads and mucous vesicles, and will aid the growing interest in engineering similar bioinspired material systems.
\end{abstract}

\maketitle

\section{Introduction}

\begin{figure}[ht]
    \centering
    \includegraphics[width=\textwidth]{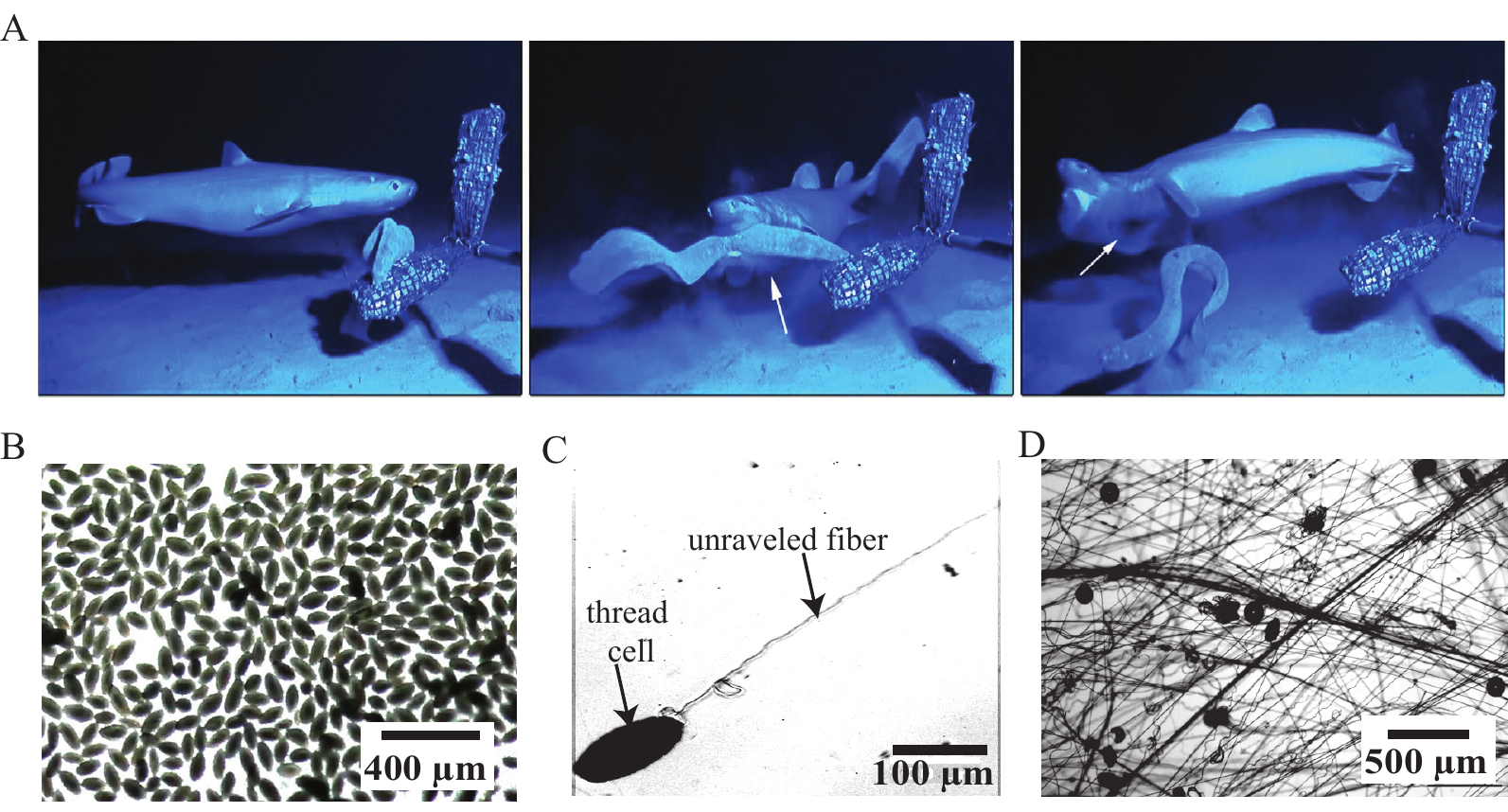}
    \caption{Slime defends hagfish against predator attacks. (A) Sequence of events during a predator attack (adapted from \citep{Zintzen2011}). On being attacked, the hagfish produces a large quantity of slime that chokes the predator. The process of secretion and slime creation took less than $0.4\,\second$. (B) Slime is formed from the secreted biomaterial, in part containing prolate-shaped thread cells. (C) A thread cell unravels under the hydrodynamic forces from the surrounding flow field and produces a micron-width fiber of length $10$--$15\,\cm$. (D) The unraveled fibers and mucous vesicles entrain a large volume of water to form a cohesive network. Details on materials and microscopy are provided in Supplementary Information (S.I.) Sec.~\ref{sec:materials}.}
    \label{fig:introduction}
\end{figure}

Marine organisms present numerous interesting examples of fluid-structure interactions that are necessary for their physiological functions such as feeding~\citep{Bishop2008,Yaniv2014}, motion~\citep{Chapman2011}, mechanosensing~\citep{Oteiza2017}, and defense~\citep{Waggett2006}. A rather remarkable and unusual example of fluid-structure interaction is the production of hagfish slime, also known as hagfish defense gel. The hagfish is an eel-shaped deep-sea creature that produces the slime when it is provoked~\citep{Downing1981}. Slime is formed from a small amount of biomaterial ejected from the hagfish's slime glands into the surrounding water~\citep{Fudge2005}. The biomaterial expands by a factor of 10,000 (by volume) into a mucus-like cohesive mass, which is hypothesized to choke predators and thus provide defense against attacks (Fig.~\ref{fig:introduction}A) \citep{Zintzen2011}.

The secreted biomaterial has two main constituents --- gland mucus cells and gland thread cells --- responsible for the mucus and fibrous component of slime, respectively~\citep{Downing1981,Fernholm1981}. In the present study we focus on thread cells, which possess a remarkable structure wherein a long filament $(10$--$16\,\cm$ in length) is efficiently packed in canonical loops into a prolate spheroid ($120$--$150\,\microm$ by $50$--$60\,\microm$)~\citep{Fernholm1981,Fudge2005}, called the skein (Fig.~\ref{fig:introduction}B). When mixed with the surrounding water, the fiber ($1$--$3\,\microm$ thread diameter) unravels from the skein  (Fig.~\ref{fig:introduction}C) and forms a fibrous network with other threads and mucous vesicles. This process occurs on timescales of a predator attack ($100$--$400\,\millisecond$), as apparent from the video evidence~\citep{Zintzen2011,Lim2006}.

While several studies have revealed the mechanical and biochemical aspects~\citep{Ewoldt2011, Winegard2014, Boni2016, Boni2018, Chaudhary2018} of slime, little is known about mechanisms involved in its rapid deployment. Newby~\citep{newby1946} postulated that the fiber is coiled under a considerable pressure and the rupture of the cell membrane allows the fiber to uncoil. However, later studies~\citep{koch1991, Lim2006, Winegard2010} have shown that convective mixing is essential for the production of fibers and slime. More recently, Bernards et al.~\citep{bernards2014} experimentally demonstrated that Pacific hagfish thread cells can unravel even in the absence of flow, potentially due to chemical release of the adhesives holding the fiber together, but the timescales observed in their work are orders of magnitude larger than physiological timescales during the attack. Therefore, the key question about the fast timescales involved in this process remains to be answered. Deeper insights into the remarkable process of slime formation will aid the development of bioinspired material systems with novel functionality, such as materials with fast autonomous expansion and deployment. Motivated by the aforementioned experimental studies, our objective in this paper is to investigate the role of viscous hydrodynamics in skein unraveling via a simple physical model, and thus supply a qualitative understanding of the unraveling process.

The key question we answer here is whether the viscous hydrodynamic unraveling alone can account for the fast unraveling timescales that are observed in physiological scenarios. We hypothesize that suction feeding in marine predators creates sufficient hydrodynamic stresses to aid in the unraveling of skeins and set up the slime network. We develop fundamental insight by considering only the simplest flow fields --- uniform flow and extensional flow.  Our modeling framework, however, generalizes to complex flow fields that occur in physiological conditions.

In Sec.~\ref{sec:experiment}, we present a simple qualitative experiment demonstrating the force-induced unraveling of a hagfish skein. This motivates the model paradigm that follows.  Section~\ref{sec:model} outlines the problem statement, and we derive the general governing equations. In Sec.~\ref{sec:skeinflow}, the equations are solved for skein unraveling in one-dimensional flows under different physically-relevant scenarios. In Sec.~\ref{sec:discussion} we discuss the results in more detail, including the influence of constitutive model parameters for the peel strength, and comment on the qualitative comparisons between the experimental studies and theoretical work.

\section{Unraveling experiment}
\label{sec:experiment}
\begin{figure}[ht]
\begin{center}
  \includegraphics[width=.98\columnwidth]{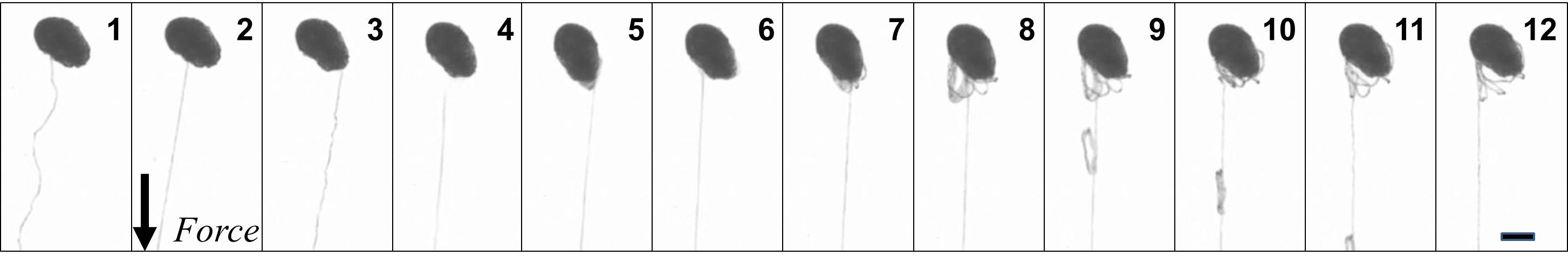}
\end{center}
\caption{Unraveling a thread skein by pulling, as viewed with brightfield microscopy. Bottom right scale bar $50\,\microm$.}
\label{fig:UnravelSequence}
\end{figure}

To motivate the mathematical modeling, we perform a simple experiment demonstrating the force-induced unraveling of thread from a skein (Fig.~\ref{fig:UnravelSequence}, see also Supplementary video). A skein, obtained from Atlantic hagfish, is held in place by weak interactions with the substrate, and a force is applied to the dangling end using a syringe tip that naturally sticks to the filament. Figure~\ref{fig:UnravelSequence} shows the unraveling skein at different time frames. Frame~1 shows the unforced and stable configuration, with no unraveling. Unraveling occurs only when a force is applied from frame 2~onward. There are events when the thread peels away in clumps, but the orderly unraveling recovers quickly. A minimum peeling force seems required to unravel the thread from the skein. A simple estimate of the minimum peeling force based on weak adhesion (van der Waals interaction) between unraveling fiber and skein gives an estimate of $0.1\,\microNewton$ (see S.I.\ Sec.~\ref{sec:force}).

\section{Problem formulation}
\label{sec:model}

\begin{figure}[t]
\begin{center}
\begin{tikzpicture}[scale=1.5]
  \draw[very thick,|-,blue] (0,0) -- (2,0)
    node[midway,above] {initial thread};
  \draw[very thick,|-,red] (2,0) -- (5,0)
    node[midway,above] {unraveled thread};
  \draw (0,-.15) node[below] {\scriptsize $\s=0$};
  \draw (2,-.15) node[below] {\scriptsize $\s=\L_0$};
  \draw (4.75,-.15) node[below] {\scriptsize $\s=\L(\t)$};
  \draw (4.75,-.45) node[below] {\scriptsize $\xv(\L,\t)=\Xv(\t)$};
  \shadedraw[ball color=red] (5.25,0) circle (.25);
  \draw (5.8,-.25) node[red] {skein};
\end{tikzpicture}
\end{center}
\caption{Simplified model of thread being drawn from a skein.  The thread has length~$\L(\t)$ with initial length~$\L(0)=\L_0$.  Here~$\s$ is the arclength material (Lagrangian) coordinate along the unraveled thread, with~$0 \le \s \le \L(\t)$.  The fixed lab (Eulerian) coordinate of the thread is~$\xv(\s,\t)$, with the thread peeling from the skein at~$\xv(\L(\t),\t) = \Xv(\t)$.}
\label{fig:drawthread}
\end{figure}
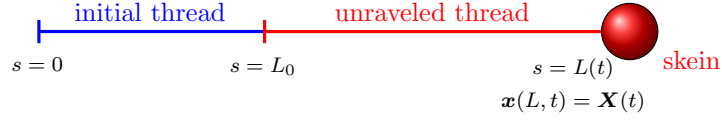

To determine if viscous hydrodynamic forces can account for fast skein unraveling, we consider a model of an inextensible slender thread unraveling from a spherical skein. The thread unravels and separates from the skein in response to a local force due to a viscous fluid flow surrounding the connected thread and skein. A schematic representation is shown in Fig.~\ref{fig:drawthread}. Here~$\xv(\s,\t)$ is the Eulerian (lab) coordinate of the centerline of the filament as a function of the Lagrangian (material) thread arclength~$\s$, $0\le\s\le\L(\t)$, with~$\L(\t)$ the time-dependent unraveled thread length.  The thread is peeling from the skein at the Eulerian point~$\xv(L(\t),\t)=\Xv(\t)$, which may depend on time if the skein is allowed to move.

\subsection{Hydrodynamic force balance}

We assume inertial effects, filament self-interactions, and external Brownian and gravitational forces to be negligible. The fluid dynamics in this situation are described by the Stokes equations. For the most general case of a thread in viscous flow, a local balance of filament forces and viscous forces (using local drag theory for a slender filament) is given by \cite{Tornberg2004}
\begin{equation}
   8\pi\dvisc\,\delta\l(\xv_\t - \uv(\xv,\t)\r)
  = -\l((1+2\delta)\,\eye + (1-2\delta)\,\shat\shat\r)\cdot\fv. \label{eq:FilamentLocalFull}
\end{equation}
Here the tangent to the thread is~$\shat$,
the dynamic viscosity is~$\dvisc$, and
\begin{equation}
  \delta=-1/\log(\eps^2\ee)>0, \quad\text{with}\quad \eps = \rr/\L,
  \label{eq:delta}
\end{equation}
are the slenderness parameter and thread aspect ratio, respectively, with~$\rr$ the thread radius.

The internal net force per unit length, $\fv$, of an inextensible filament is expressed by the Euler--Bernoulli bending theory for an elastic beam, and has both tensile and bending components:
\begin{equation}
  \fv(\s)
  = -(\T\,\xv_\s)_\s + \E\,\xv_{\s\s\s\s}\,,
  \qquad
  \lvert\xv_\s\rvert=1.
  \label{eq:fv}
\end{equation}
Here $\E$ is the bending modulus of the thread, $\T(\s,\t)$ is the tension in the filament, and each subscript~$\s$ denotes one derivative, e.g.~$\xv_s = \nofrac{\partial \xv}{\partial s}$.  The inextensibility condition is~$\lvert\xv_\s\rvert=1$, so~$\s$ and distance along the thread must always coincide.

In the spirit of rheology, we consider the response to simple flows to isolate key features of the complex behavior, obtain analytical results, and gain an understanding of the unraveling process.  We only consider cases with zero curvature, $\xv_{\s\s}=0$, immersed in one-dimensional flow fields, with the thread aligned with the flow streamlines. Equation~\ref{eq:FilamentLocalFull} then reduces to a one-dimensional statement that the component of internal net filament force per unit length $\fc(\s)$ along the streamline (taken as the~$\xc$ direction) is equal to the local viscous drag per unit length:
\begin{equation}
   4\pi\dvisc\,\delta\l(\xc_\t - \uc(\xc,\t)\r)
   = -\fc.
\end{equation}
Then the one-dimesional form of Eq.~\eqref{eq:fv} with~$\xc_{\s\s\s\s}=0$ and the inextensibility condition~$\xc_\s=1$ gives~$\fc = -\T_\s$, so that
\begin{equation}
  \T_\s =
  4\pi\dvisc\,\delta\l(\xc_\t - \uc(\xc,\t)\r).
  \label{eq:Ts}
\end{equation}
With~$\xc_\s=1$ and~$x(\L,\t)=\X(\t)$ we have~$\xc = \X - \L + \s$, where~$\X$ is the skein position, and thus~$x_\t = \dot\X - \dot\L$. We integrate~\eqref{eq:Ts} from~$\s=0$ to~$\L$ to find
\begin{equation}
  \T|_{\s=\L} - \T|_{\s=0} =
  4\pi\dvisc\L\,\delta\l(\dot\X-\dot\L
  - \frac{1}{\L}\int_0^\L \uc(\X - \L + \s,\t)\dint\s\r).
\end{equation}
We then change the integration variable to~$\xc = \X - \L + \s$, and finally
obtain
\begin{equation}
  \TL - \Tz
  = -4\pi\dvisc\L\,\delta\l(\dot\L - \dot\X + \bar\uc(\L,\X,\t)\r)
   \label{eq:netdrag}
\end{equation}
where~$\TL = \T|_{\s=\L}$, $\Tz = \T|_{\s=0}$, and~$\bar\uc(\L,\X,\t)$ is the
average velocity on the filament:
\begin{equation}
  \bar\uc(\L,\X,\t) \ldef \frac{1}{\L} \int_{\X-\L}^{\X}\uc(\xc,\t)\dint\xc\,.
  \label{eq:ubar}
\end{equation}
Equation~\eqref{eq:netdrag} expresses the balance between the tension forces at the end of the thread and the drag force on the thread.  We shall use this equation to derive a peeling formula for different thread-skein configurations in Sec.~\ref{sec:skeinflow}.  But first we need to examine how the thread will peel from the skein to unravel.

\subsection{Unraveling from the skein}
\label{sec:unravelskein}

The relationship between $\R$ and~$\L$, respectively the radius of the spherical skein and the length of the unraveled thread, is described by volume conservation
\begin{equation}
  \frac{d}{d\t}\l(\tfrac43\pi\eta\R^3 + \pi \rr^2 \L\r) = 0
  \quad\Longrightarrow\quad
  \dot\L = -4\eta\R^2\dot\R/\rr^2.
  \label{eq:dRdL}
\end{equation}
Here~$\rr$ is the thread radius and~$0 < \eta \le 1$ is the packing fraction of
thread into the spherical skein, assumed independent of~$\R$.  (In this section we keep the packing fraction as a variable, but in all later numerical simulations we take~$\eta=1$, since the skein is fairly tightly packed.)  Explicitly, we
have
\begin{equation}
  \R^3
  =
  \R_0^3 - \tfrac34 (\L-\L_0)\rr^2/\eta
  \label{eq:RfromL}
\end{equation}
with $\R_0$ the initial skein radius and $\L_0$ the initial unraveled length.  A convenient way of relating~$\R$ and~$\L$ is
\begin{equation}
  \R =
  \R_0\l(\frac{\Lmax - \L}{\Lmax - \L_0}\r)^{1/3},
  \qquad
  \Lmax \ldef \L_0 + \tfrac43\eta\R_0^3/\rr^2
  \label{eq:Lmax}
\end{equation}
where~$\Lmax$ is the total length of thread that can be extracted and $\L_0$ is the initial unraveled length.

Next, we use a modified form of the work-energy theorem \citep{Hong1995} to describe the unraveling dynamics,
\begin{equation}
  \dotEner = (\TL - \Fp(\V))\,\V,
  \qquad
  \V = \dot\L\,,
  \label{eq:WE}
\end{equation}
where~$\dotEner$ is the rate of change in total energy of the system, $\TL$ is
the net force (given by Eq.~\eqref{eq:netdrag}) drawing out the thread at a peeling velocity~$\V$, and~$\Fp(\V)$ is a velocity-dependent peeling force acting at the peeling site. Neglecting the inertia and changes to the elastic energy of the peeling thread gives
\begin{equation}
  \TL = \Fp(\V),
  \qquad
  \V = \dot\L\,.
  \label{eq:TmF0}
\end{equation}

A natural dimensionless quantity that will determine the dynamics of the unraveling process is given by the ratio of the net viscous drag force on the thread and the resisting peel force, each of which depends on a characteristic velocity $U$:
\begin{equation}
  \wp \ldef \nofrac{\Fd(\U)}{\Fp(\U)}\,.
  \label{eq:forceratio}
\end{equation}

The functional form of the peeling force, $\Fp(\V)$, in general is dependent on parameters such as the chemistry of peeling surfaces, velocity of peeling, etc. In the absence of a known functional form for hagfish thread peeling, we use a simple constitutive form of peeling force that includes a wide range of behavior, given by
\begin{equation}
  \Fp(\V)=\FVconst\V^\FVpow,\qquad 0\le\FVpow\le1,
  \label{eq:FpV}
\end{equation}
for constant~$\FVconst>0$ and~$\FVpow$. Such a power-law form of peeling force has been observed in several engineered and biological systems \citep{Cortet2007, mohammed2016, liu2018, dalbe2014, wang2018}. Several other parametric forms of velocity-dependent peeling force exist that are functionally more complex \citep{De2004,Griffin2004}. However, to obtain simple and insightful solutions, we use the power-law form defined above. The form \eqref{eq:forceratio} allows for the limiting case $\FVpow=0$, a constant peeling force, e.g.~to simply counteract van der Waals attractions at the peel site.

For~$\FVpow > 0$, we can rearrange equation~\eqref{eq:TmF0} for the velocity, $\V = \dot\L =(\TL/\FVconst)^{1/\FVpow}$.  Using~\eqref{eq:dRdL} we can then obtain a solution for the case where the tension at the peeling point, \TL, is constant:
\begin{equation}
  \tfrac43(\R_0^3-\R^3)
  = \l(\TL/\FVconst\r)^{1/\FVpow}\rr^2\t/\eta\,,
  \label{eq:RsolTconst}
\end{equation}
where~$\R_0=\R(0)$.  From~\eqref{eq:RsolTconst} we can easily
extract the `depletion time' or `full-unraveling time'~$\tdep$ by
setting~\hbox{$\R=0$}:
\begin{equation}
  \tdep = \frac{4\eta\R_0^3}{3\rr^2}\l(\TL/\FVconst\r)^{1/\FVpow}.
\end{equation}
In the next section we compute this timescale when the thread cells are subjected to different hydrodynamic flow scenarios, which cause different time histories of tension,~$\TL(t)$.

\section{Thread cell in one-dimensional flow}
\label{sec:skeinflow}

Having described the unraveling dynamics in Sec.~\ref{sec:unravelskein} for the case of constant tension, \TL, we now consider a thread cell (skein) in a hydrodynamic flow where generally $\TL$ varies in time as the thread-skein geometry changes during unraveling. To simplify the problem we assume an incompressible flow of the form
\begin{equation}
  \uv(\xc,\yc,\t)=(\uc(\xc,\t),-\yc\uc_\xc(\xc,\t)).
  \label{eq:horizflow}
\end{equation}
The thread will be assumed to lie along the~$\xc$ axis. We solve for the depletion time for four relevant cases: Pinned thread in uniform flow (Sec.~\ref{sec:pinnedthread}); Pinned skein in uniform flow (Sec.~\ref{sec:pinnedskein}); Free skein and thread in extensional flow (Sec.~\ref{sec:freeskeinthread}); and Free skein splitting into two smaller skeins  in extensional flow (Sec.~\ref{sec:twoskeins}).

\subsection{Pinned thread}
\label{sec:pinnedthread}

The simplest case to consider is the thread pinned at~$\s=0$ in Fig.~\ref{fig:drawthread}, with a uniform flow to the right, $\uc(\xc,\t)=\U$. This situation can arise in a controlled experiment if the thread is pinned down, or in the physiological unraveling process if the end of the thread is caught in the network of other threads, or stuck on the mouth of a predator.

The tension in the thread at~$\s=\L$ balances the Stokes drag on the skein of radius~$\R$, $\TL = \T(\L(\t),\t) = 6\pi\dvisc\R\,(\uc(\L,\t) - \dot\L)$. Using \eqref{eq:TmF0} and \eqref{eq:FpV}, we obtain the governing equation for unraveling as
\begin{equation}
  (\dot\L)^\FVpow
  = 6\pi\dvisc\FVconst^{-1}\R(\L)\,(\uc(\L,\t) - \dot\L).
  \label{eq:pinnedthread}
\end{equation}
From \eqref{eq:dRdL}, since $\dot\L>0$ (the thread never `re-spools'), the unspooling speed satisfies~\hbox{$\dot\L \le \uc(\L,\t)$}, i.e.~the thread cannot unspool faster than the ambient flow speed.  The radius~$\R(\L)$ is given by~\eqref{eq:RfromL}.

We nondimensionalize \eqref{eq:pinnedthread} using a characteristic length scale $\R_0$ and flow speed $\U$ which gives
\begin{equation}
    (\dot\L^*)^\FVpow
  = \wp\,\R^*(\L^*)\,(\uc^*(\L^*,\t^*) - \dot\L^*)
  \label{eq:pinnedthreadnondimensional}
\end{equation}
where $\dot\L^* = {\dot{\L}}/{\U}$, $\R^* = {\R}/{\R_0}$ and $\uc^*(\L^*,\t^*) = {\uc(\L,\t)}/{\U}$ are the nondimensional unraveling rate, skein radius, and flow rate, respectively. The nondimensional timescale naturally results from these choices as $\t^* = \nofrac{t}{(\R_0/\U)}$. The dimensionless quantity $\wp$ on the right hand side of \eqref{eq:pinnedthreadnondimensional} is given by
\begin{equation} \label{eq:Pcase1}
  \wp  =
  \frac{6\pi\dvisc\R_0\,\U}{\FVconst\U^\FVpow}
  =
  6\pi\dvisc\R_0\,\U^{1-m}\FVconst^{-1}\,.
\end{equation}
This is the ratio of characteristic drag to peeling force, as defined in \eqref{eq:forceratio}. If~$\wp$ is large (e.g. zero resistance to peeling), then~\eqref{eq:pinnedthread}
implies~$\dot\L \approx \uc(\L,\t)$, that is, in this drag-dominated limit the drag force so easily unravels the skein that it advects with the local flow velocity. In the opposite limit of small~$\wp $, we get~$\dot\L \approx 0$ and the skein cannot unravel. Hence, we require $\wp \gg 1$ for a fast unravel time.

To achieve the criterion $\wp \gg 1$, at a flow of speed~$\U = 1\,\meter/\second$ and a skein of initial radius~$\R_0=50\,\microm$, we  require the peeling resistance at this velocity to satisfy $\Fp(1\,\meter/\second)\ll 1.4 \times 10^{-6}\,\Newton$. The estimated van der Waals peeling force is much lower than this threshold, $F_{\text{vdW}}\sim10\,\microNewton$. At such a flow speed a skein containing~$16.7\,\cm$ of thread (an upper bound physiological value) will unravel affinely (kinematically matching the flow speed) in roughly~$167\,\millisecond$. This lower bound estimate is commensurate with the rapidity with which hagfish slime is created ($100$--$400\,\millisecond$).

\begin{figure}
\begin{center}
  \includegraphics[width=.7\columnwidth]{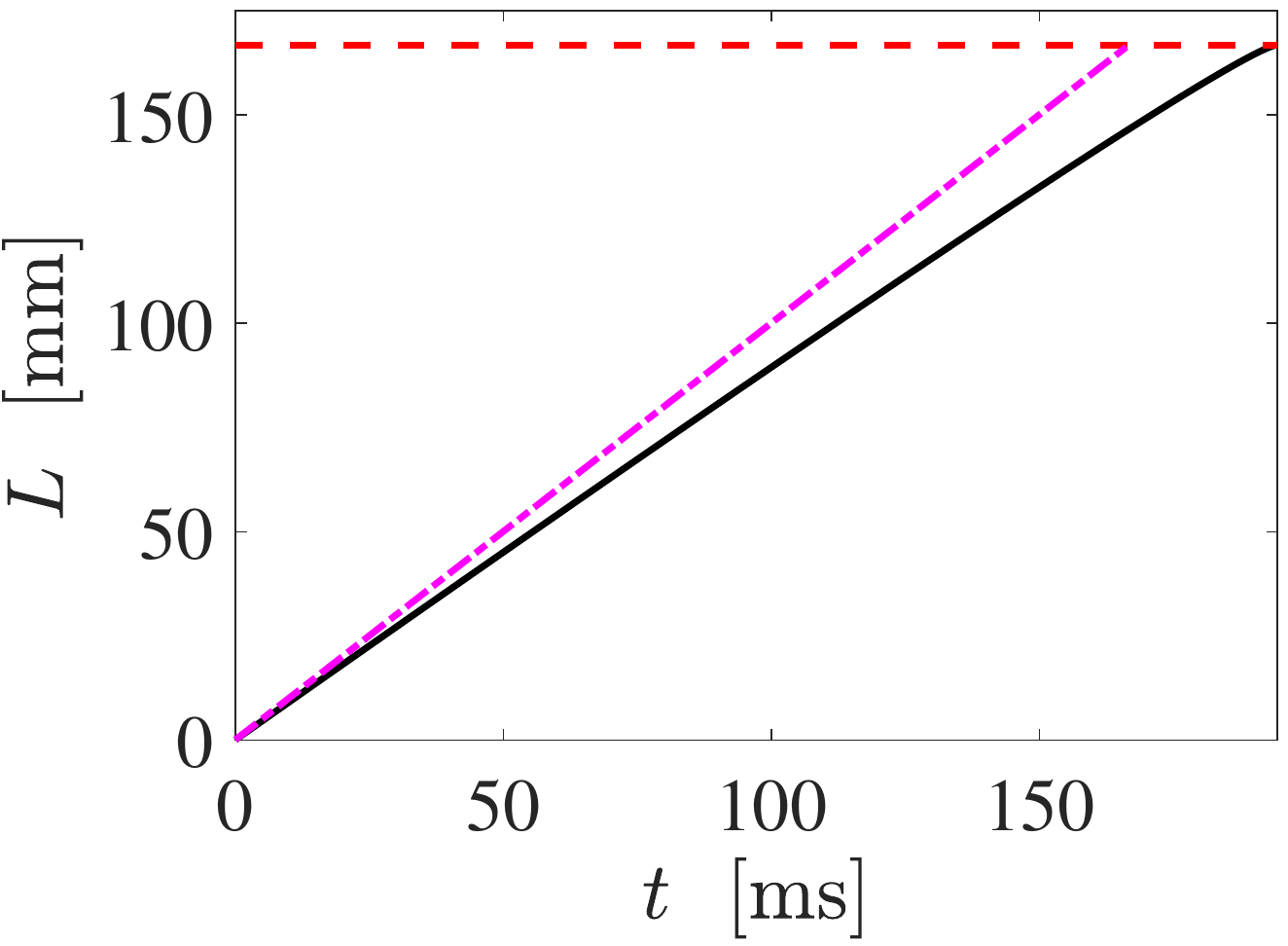}
\end{center}
\caption{Numerical solution (solid line) of~\eqref{eq:pinnedthread} for the
  parameter values $\R_0=50 \,\microm$, $\L_0=2\R_0$, $\wp = 10$,
  $\FVpow=1/2$, $\U = 1\,\meter/\second$.  The \textbf{-- ---} line is the
  upper bound~$\L = \L_0 + \U\t$.  The horizontal dashed line is
  at~$\L=\Lmax$, when the skein is fully unraveled.  Even for such a moderate
  force ratio~$\wp=10$ the thread unravels almost as fast as the upper bound.}
\label{fig:Lsolve_pinned_thread}
\end{figure}

In Fig.~\ref{fig:Lsolve_pinned_thread} we show a numerical solution
of~\eqref{eq:pinnedthread} with a uniform flow for some typical physical
parameters values, and assuming a moderately-large force ratio~$\wp = 10$.
(Equation~\eqref{eq:pinnedthread} is an implicit relation for~$\dot\L$ which
must be solved numerically at every time step; it is a Differential-Algebraic Equation rather than a simple ODE~\cite{De2004}.)  For these parameters, the kinematic lower bound on the depletion time is~$\Lmax/\U \approx 167\,\millisecond$, and the numerical value is~$\tdep \approx 194\,\millisecond$.

There is a mathematical oddity where the skein might not get depleted in finite time, depending on the exponent~$\FVpow$. To see this, consider a skein close to depletion, $\L = \Lmax - \U\tau$, where~$\tau>0$ is small. The equation for~$\tau$ is
\begin{equation}
  (-\dot\tau)^\FVpow
  = \wp\l(\frac{\U\tau}{\Lmax - \L_0}\r)^{1/3}
  \,(1 + \dot\tau),
  \qquad
  \dot\tau < 0.
\end{equation}
Since~$\tau$ is small and we expect the thread to be drawn out slowly as
it is almost exhausted, we take~$1 + \dot\tau \approx 1$.  Hence, we have the
approximate form
\begin{equation}
  (-\dot\tau)^\FVpow
  \approx \text{C}^\FVpow\,\tau^{1/3},
  \qquad
  \text{C}^\FVpow \ldef \wp\,(\U/(\Lmax - \L_0))^{1/3}
  \label{eq:onethird}
\end{equation}
for some constant~$\text{C} > 0$, with solution
\begin{equation}
  \tau(\t)
  \approx
  \l[\tau_0^{1-\frac{1}{3\FVpow}}
  - \l(1-\tfrac{1}{3\FVpow}\r)\text{C}\,\t\r]^{\frac{3\FVpow}{3\FVpow-1}}
  .
  \label{eq:ellsol}
\end{equation}
The behavior of this solution as the skein is almost depleted depends
on~$\FVpow$.  For~$\FVpow>1/3$, the exponent~$3\FVpow/(3\FVpow-1)$
in~\eqref{eq:ellsol} is greater than one, so~$\tau(\t) \rightarrow 0$ as~$\t$
approaches the depletion time, with~$\tau'(\tdep)=0$ so that~$\L(\t)$ has
slope zero when the skein is depleted (as can be seen at the very end in
Fig.~\ref{fig:Lsolve_pinned_thread}).  We can thus rewrite~\eqref{eq:ellsol}
as
\begin{equation}
  \tau(\t)
  \approx
  \l[\l(1-\tfrac{1}{3\FVpow}\r)\text{C}\,(\tdep - \t)\r]^{\frac{3\FVpow}{3\FVpow-1}}
  ,
  \qquad
  \FVpow > 1/3,
  \quad
  \t \nearrow \tdep.
\end{equation}

For~$\FVpow < 1/3$, the exponent~$3\FVpow/(3\FVpow-1)$ is negative, but the
factor~$1-\tfrac{1}{3\FVpow}$ inside the brackets is also negative, so that
$\tau(\t)$ asymptotes to zero as~$\t \rightarrow \infty$ and the skein never
gets fully depleted.  In that case we write~\eqref{eq:ellsol} as
\begin{equation}
  \tau(\t)
  \approx
  \l[\l(\tfrac{1}{3\FVpow}-1\r)\text{C}\,\t\r]^{-\frac{3\FVpow}{1-3\FVpow}}
  ,
  \qquad
  \FVpow < 1/3,
  \quad
  \t \rightarrow \infty.
\end{equation}
Physically, for~$\FVpow < 1/3$ the drag force
($\sim\tau^{1/3}$) is decreasing faster than the peeling force
($\sim(\dot\tau)^\FVpow$).

In practice, it is difficult to see the difference
between~$\FVpow \lessgtr 1/3$ numerically.  The thread appears to get depleted
even for~$\FVpow < 1/3$ because of limited numerical precision as~$\L$
approaches $\Lmax$.  The symptom of a problem is that the depletion time
starts depending on the numerical resolution for~$\FVpow < 1/3$.  Of course,
the skeins in the hagfish slime do not need to get fully depleted to create
the gel, so a power~$\FVpow < 1/3$ is still applicable. When comparing the different flow scenarios we will explore a range of $\FVpow$ and define an ``effective deployment'' time~$\tdepfifty$, when 50\% of the thread length is unraveled.

\subsection{Pinned skein}
\label{sec:pinnedskein}

When the skein is pinned and the thread is free at the other end, the tension arises from hydrodynamic drag on the thread. Consider a free thread ending at~$\s=0$ and a pinned skein at~$\s=\L(\t)$ (Fig.~\ref{fig:drawthread}), so that the Eulerian skein position~$\X$ is fixed and is thus not a function of time.  Unlike the pinned thread case in Sec.~\ref{sec:pinnedthread}, where a shrinking skein led to a decreasing drag, here the tension \emph{increases} with time as the extended thread provides more drag.

We formulate the problem by imposing boundary conditions at the free end, $\Tz = \T(0,\t) = 0$, and pinned end, $\xc(\L(\t),\t) = \X$.  From~\eqref{eq:netdrag} with~$\Tz=\dot\X=0$ and~$\TL = \FVconst\,(\dot\L)^\FVpow$, the equation for the growth of the thread is
\begin{equation}
  (\dot\L)^\FVpow
  = -4\pi\dvisc\,\FVconst^{-1}\L\,\delta(\L)\,
  \bigl(
  \dot\L + \bar\uc(\L,\X,\t)
  \bigr).
  \label{eq:pinnedskein}
\end{equation}
The slenderness parameter~$\delta$ depends on~$\L$ through its definition~\eqref{eq:delta}. Because the thread extends to the left in Fig.~\ref{fig:drawthread}, we must have~$\bar\uc(\L,\X,\t) < 0$ to avoid unphysical respooling.  The pinned thread equation~\eqref{eq:pinnedthread} and the pinned skein equation~\eqref{eq:pinnedskein} have a very similar form, though the drag in the former ($\sim\R(\L)$) decreases with~$\L$ and that in the latter ($\sim\L\delta(\L)$) increases with~$\L$.

Using a characteristic velocity~$\U$ and a characteristic length scale $\L_0$, we obtain the nondimensional form of \eqref{eq:pinnedskein} as
\begin{equation}
  (\dot\L^*)^\FVpow   = -\wp\L^* \delta(\L^*) (\bar\uc^*(\L^*,\X^*,\t^*)
  + \dot\L^*\bigr),
  \label{eq:pinnedskeinnondimensional}
\end{equation}
where $\dot\L^* = {\dot{\L}}/{\U}$, $\L^* = {\L}/{\L_0}$, and $\uc^*(\L^*,\t^*) = {\uc(\L,\t)}/{\U}$ are nondimensional unraveling rate, unraveled length and flow rate, respectively. The natural dimensionless force ratio~\eqref{eq:forceratio} is
\begin{equation}
  \wp = 4\pi\dvisc\,\L_0\,\U^{1-\FVpow}\,\FVconst^{-1}.
  \label{eq:Uscale}
\end{equation}
This differs from $\wp$ in \eqref{eq:Pcase1} by replacing $R_0$ with $\L_0$. It is sensible in this pinned skein case to use the initial thread length~$\L_0$, since drag on the thread controls the unraveling rate.

\begin{figure}
\begin{center}
  \includegraphics[width=.7\columnwidth]{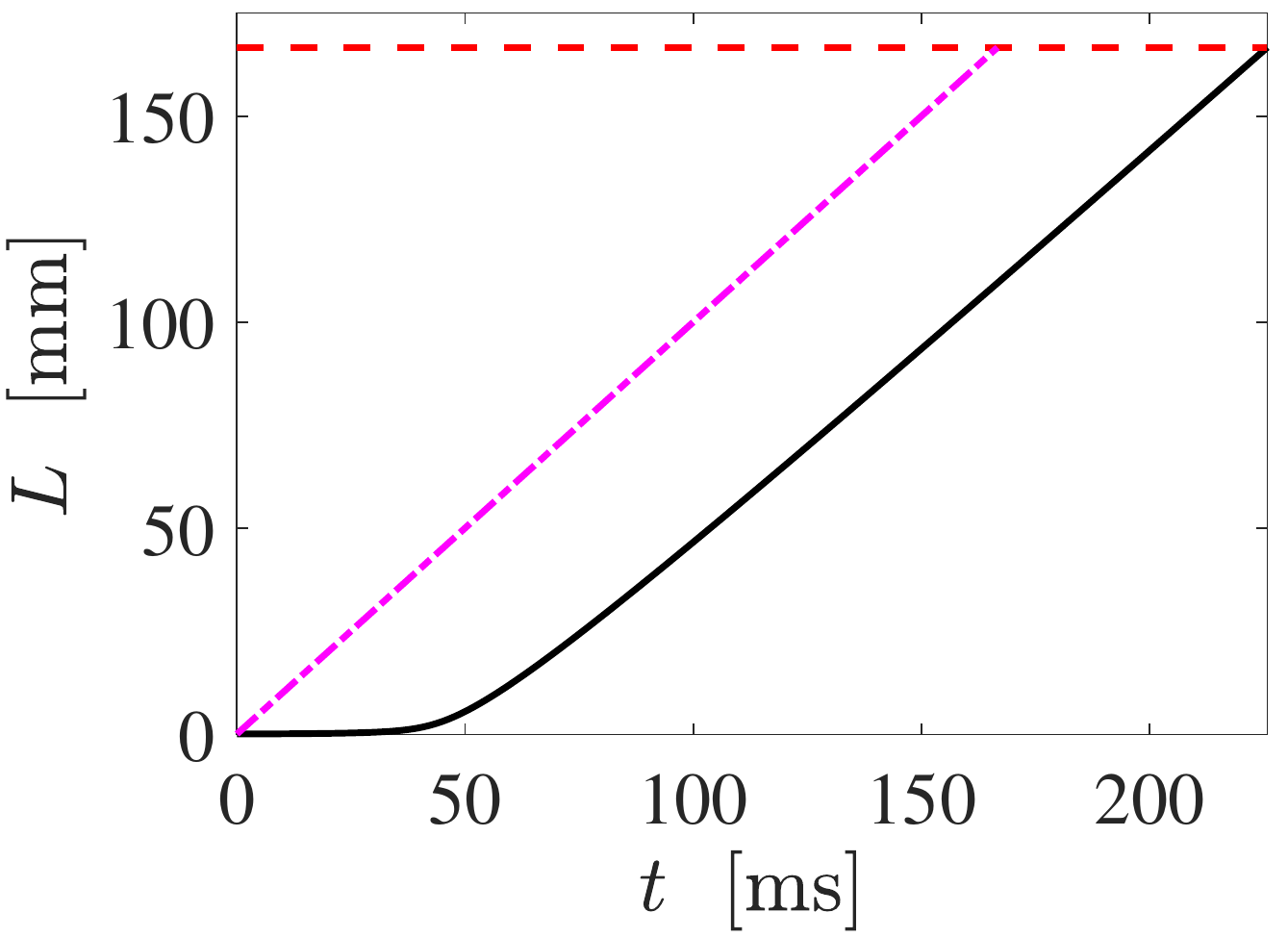}
\end{center}
\caption{Numerical solution (solid line) of~\eqref{eq:pinnedskein} for the
  parameter values $\R_0=50 \,\microm$, $\L_0=2\R_0$, $\wp = 1/2$,
  $\FVpow=1/2$, $\U = 1\,\meter/\second$.  The \textbf{-- ---} line is the
  upper bound~$\L = \L_0 + \U\t$.  The horizontal dashed line is
  at~$\L=\Lmax$, when the skein is fully unraveled.  Even for such a small
  force ratio~$\wp$ the thread unravels almost as fast as the upper bound.}
\label{fig:Lsolve_pinned_skein}
\end{figure}

Figure~\ref{fig:Lsolve_pinned_skein} shows a numerical solution of~\eqref{eq:pinnedskein} for our reference parameter values using a constant velocity field, $\uc(\L,\X,\t)=-\U=-1\,\meter/\second$. As before, the lower bound on the depletion time is~$\Lmax/\U \approx 167\,\millisecond$, and now the numerical value is~$\tdep \approx 226\,\millisecond$. This is slower than what we observed in the pinned thread case ($\tdep \approx 194\,\millisecond$), but here we are using the much smaller force ratio~$\wp = 1/2$. This shows that the pinned skein case can unravel almost as fast as the lower bound for a much smaller value of~$\wp$, since the drag on the thread increases with~$\L$, as reflected by the accelerating speed $\dot\L$ in Fig.~\ref{fig:Lsolve_pinned_skein}.  This is in contrast to the deceleration in Fig.~\ref{fig:Lsolve_pinned_thread} for the pinned thread, where drag decreases as the skein radius diminishes.

\subsection{Free skein and thread}
\label{sec:freeskeinthread}
In the previous two cases, we took either the thread or skein to be pinned; here we consider the case where neither is pinned, and both are free to move with the flow.  The force at the peeling point~$\xc(\L(\t),\t)=\X(\t)$ is then determined by the balance of two forces: Stokes drag on the spherical skein, $F_1 = 6\pi\dvisc\R\,(\uc(\X,\t) - \dot\X)$, and drag on the thread, $F_2 = -4\pi\dvisc\L\delta\,(\dot\L - \dot\X + \bar\uc(\L,\X,\t))$.  The latter is obtained from~\eqref{eq:netdrag} with~$\TL=F_2$ and~$\Tz=0$.  Since both the skein and thread are free and we have neglected inertia, $F_1+F_2=0$, which we can use to solve for $\dot\X$, the velocity of the peeling point in the Eulerian (lab) frame. Coupling this with the peel force constitutive model~\eqref{eq:TmF0}--\eqref{eq:FpV}, the unspooling rate equation is then~$\FVconst(\dot\L)^\FVpow = F_1 = -F_2$.  The dynamics of this scenario are governed by the system
\begin{subequations}
\begin{align}
  (\dot \L)^\FVpow
  &=
  -\frac{12\pi\dvisc\FVconst^{-1}\R\L\delta}{2\L\delta + 3\R}\,
  (\dot\L + \bar\uc(\L,\X,\t) - \uc(\X,\t)),
  \label{eq:dLfreefree} \\
  \dot\X
  &=
  \frac{2\L\delta}{2\L\delta + 3\R}\,
  (\dot\L + \bar\uc(\L,\X,\t))
  +
  \frac{3\R}{2\L\delta + 3\R}\,
  \uc(\X,\t)
  ,
  \label{eq:dXfreefree}
\end{align}
\label{eq:freefree}%
\end{subequations}
where~$\bar\uc(\L,\X,\t)$ is the thread-averaged velocity~\eqref{eq:ubar}. The velocity~\eqref{eq:dXfreefree} for the thread-skein system is the average of a velocity~$\dot\L + \bar\uc(\L,\X,\t)$ arising from drag on the thread and a velocity~$\uc(\X,\t)$ arising from drag on the skein, weighed by the relative strength of the drags.

The difference~$\bar\uc(\L,\X,\t) - \uc(\X,\t)$ that appears in~\eqref{eq:dLfreefree} implies that adding a constant to the velocity field does not change the unspooling dynamics, as expected since the thread-skein system is freely advected by the flow, and only relative velocities generate drag.  Hence, unlike our previous two cases in Secs.~\ref{sec:pinnedthread}--\ref{sec:pinnedskein}, a spatially-varying flow field is required for unraveling. For a linear velocity field~$\uc(\xc,\t) = \lambda\xc$, i.e.~uniaxial extensional flow with extensional strain rate $\lambda$, we have~$\uc(\X,\t) - \bar\uc(\L,\X,\t) = \tfrac12\lambda\,\L$ independent of~$\X$, so that we can solve the~$\dot\L$ equation~\eqref{eq:dLfreefree} by itself:
\begin{equation}
  (\dot \L)^\FVpow
  =
  \frac{6\pi\dvisc\FVconst^{-1}\R\L}{\L + (\nofrac{3\R}{2\delta})}\,
  (\tfrac12\lambda\,\L - \dot\L).
  \label{eq:dLfreefreestrain}
\end{equation}
The mass conservation equation~\eqref{eq:RfromL} then relates~$\R$ to~$\L$,
and the slenderness parameter~\eqref{eq:delta} relates~$\delta$ to~$\L$.

To define a characteristic length scale for this problem, should we use $\R_0$ or $\L_0$ as a length scale?  Both are important for the unraveling process to start quickly, but typically~$\L_0$ is a bit larger than~$\R_0$.  A compromise is to use~$\R_0$ as the viscous drag length scale and~$\U = \lambda\L_0$ as the velocity scale.  The choice of~$\R_0$ emphasizes the magnitude of the drag on the skein, and~$\lambda\L_0$ reflects the amplitude of velocity gradients over the longer length~$\L_0$.  We thus obtain the dimensionless form of \eqref{eq:dLfreefreestrain} as
\begin{equation}
(\dot \L^*)^\FVpow =
  \wp\,\frac{\R^*\L^*}{\L^* + (\nofrac{3\R^*}{2\delta})}\,
  (\tfrac12\L^* - \dot\L^*)
\end{equation}
where $\dot\L^* = {\dot{\L}}/{\lambda \L_0}$, $\R^* = {\R}/{\R_0}$, and $\L^* = {\L}/{\R_0}$ are nondimensional unraveling rate, skein radius and unraveled length, respectively.  The natural dimensionless number in this case is
\begin{equation}
  \wp = \frac{6\pi\dvisc\,\R_0\,\U}{\FVconst\,\U^{\FVpow}}
  = 6\pi\dvisc\,\R_0\,(\lambda\,\L_0)^{1-\FVpow}\,\FVconst^{-1}.
  \label{eq:Pcase3}
\end{equation}

Assuming as before that~$\dot\L\ge0$ (the thread doesn't `re-spool'), the right-hand side of~\eqref{eq:dLfreefreestrain} implies~$\dot\L \le \tfrac12\lambda\,\L$, which gives the constraint that $  \L(\t) \le \L_0\,\ee^{\tfrac12\lambda\t}$.  This constraint is the kinematic limit where the thread extends at a rate dictated by the strain rate in the flow.  This implies that the depletion time satisfies
\begin{equation}
  \tdep \ge 2\lambda^{-1}\log(\Lmax/\L_0).
  \label{eq:tdepexpbound}
\end{equation}
In the two pinned cases we considered before, the lower bound on the depletion time was of the form~$\tdep \ge \Lmax/\U$, \emph{independent of~$\L_0$}.  The lower bound~\eqref{eq:tdepexpbound} depends explicitly on the ratio~$\Lmax/\L_0$, so a very short initial thread length will take a long time to unravel, even if~$\wp$ is large.

\begin{figure}
\begin{center}
  \includegraphics[width=.7\columnwidth]{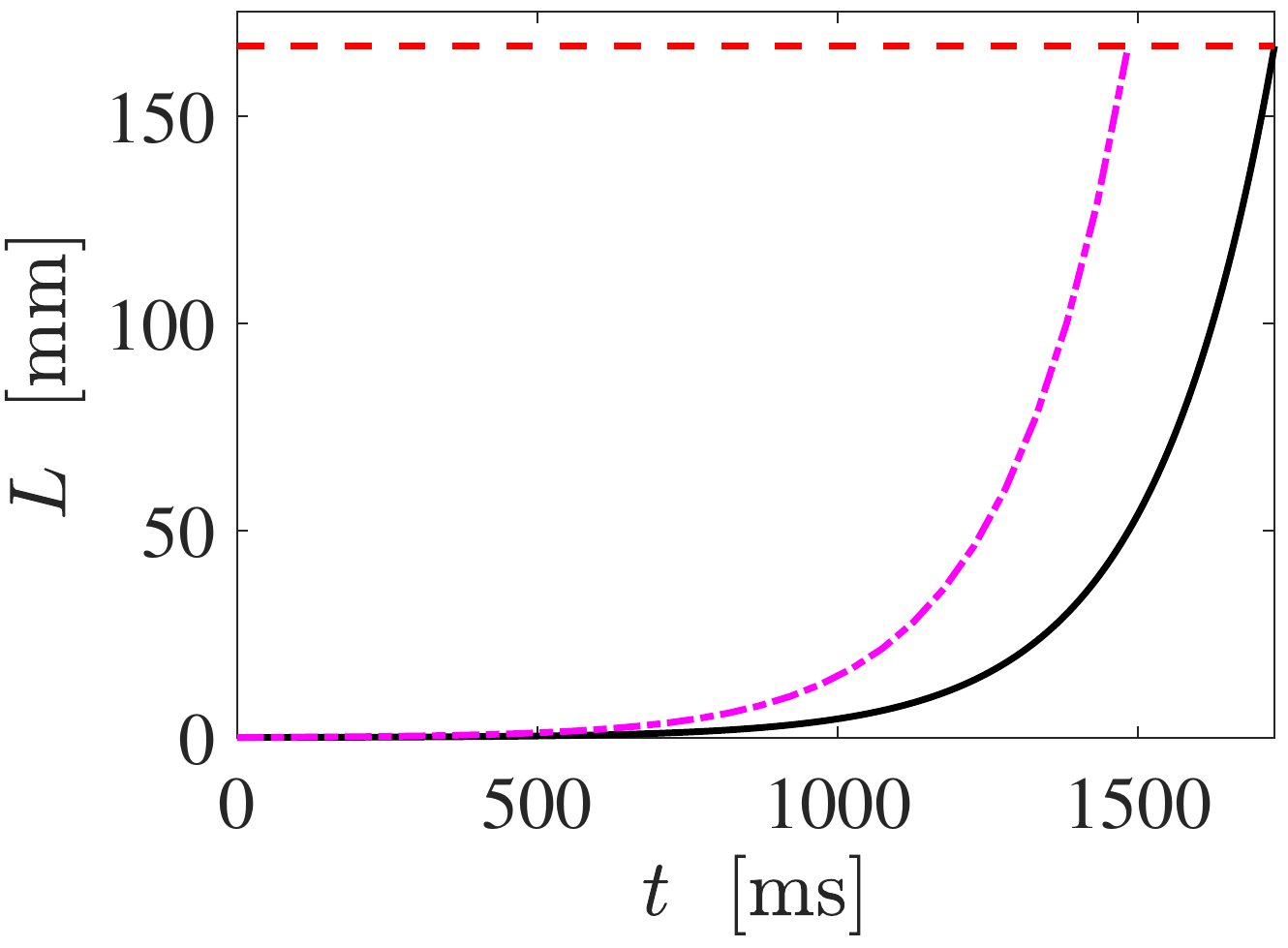}
\end{center}
\caption{Numerical solution (solid line) of~\eqref{eq:dLfreefreestrain} for the parameter values $\R_0=50 \,\microm$, $\L_0=2\R_0$, $\wp = 10$, $\FVpow=1/2$, $\lambda = 10\,\second^{-1}$.  The \textbf{-- ---} line is the upper bound~$\L_0\exp(\tfrac12\lambda\t)$.  The horizontal dashed line is at~$\L_1=\Lmax$, when the skein is fully unraveled.}
\label{fig:Lsolve_freefree}
\end{figure}

When the thread is almost depleted, the unspooling rate decreases due to the factor of~$\R$ in~\eqref{eq:dLfreefreestrain}.  To see this explicitly, put~$\L = \Lmax - \U\tau$ in~\eqref{eq:dLfreefreestrain} and assume~$\tau$ and~$\dot\tau$ are small:
\begin{equation}
  (-\dot\tau)^\FVpow
  \approx
  \tfrac12\wp\l(\frac{\U\tau}{\Lmax - \L_0}\r)^{1/3}\,\frac{\Lmax}{\L_0},
  \qquad \tau \ll 1.
\end{equation}
This is exactly the same form as~\eqref{eq:onethird}, with a different constant~$\text{C}$.  We conclude that once again the criterion for finite-time complete unraveling is~$\FVpow > 1/3$, as it was for the pinned thread case (Sec.~\ref{sec:pinnedthread}).  But as before this is not very physically consequential, as it only applies to the last phase of unspooling when the skein is almost completely unraveled.

Figure~\ref{fig:Lsolve_freefree} shows a numerical solution of~\eqref{eq:dLfreefreestrain} for our reference parameter values and with a strain rate~$\lambda = 10\,\second^{-1}$ for~$\wp = 10$.  (We choose~$\lambda$ such that~$\lambda\Lmax$ is of the same order of magnitude as~$\U=1\,\meter/\second$ in the pinned cases.)  The lower bound~\eqref{eq:tdepexpbound} on the depletion time is $1.48\,\second$, and the numerical value is~$\tdep \approx 1.73\,\second$.  This is slower than what we observed in the two pinned cases ($\tdep < 1\,\second$), due to the factor~$2\log(\Lmax/\L_0) \approx 14.8$.  The slowdown due to the short initial thread length is thus considerable in this case.  A longer initial length or a higher strain rate would be needed to make the times comparable.

\subsection{Two free skeins (skein splitting)}
\label{sec:twoskeins}

\begin{figure}
\begin{center}
\begin{tikzpicture}[scale=1.5]
  \draw[very thick,blue] (-1,0) -- (1,0);
  \draw[very thick,blue] (0,-.1) -- (0,.1)
    node[above] {initial thread};
  \draw[very thick,|-,red] (1,0) -- (3.5,0)
    node[midway,above] {unraveled thread 1};
  \draw[very thick,|-,red] (-1,0) -- (-3.5,0)
    node[midway,above] {unraveled thread 2};
  \draw (0,-.15) node[below] {\scriptsize $\s=0$};
  \draw (0,-.45) node[below] {\scriptsize $\xc(0,\t)=\X(\t)$};
  \draw (1,-.15) node[below] {\scriptsize $\L_0/2$};
  \draw (3.45,-.15) node[below] {\scriptsize $\L_1$};
  \draw (-1,-.15) node[below] {\scriptsize $-\L_0/2$};
  \draw (-3.45,-.15) node[below] {\scriptsize $-\L_2$};
  \shadedraw[ball color=red] (3.75,0) circle (.25);
  \draw (4.3,-.25) node[red] {skein 1};
  \shadedraw[ball color=red] (-3.75,0) circle (.25);
  \draw (-4.38,-.25) node[red] {skein 2};
\end{tikzpicture}
\end{center}
\caption{Thread being drawn from two skeins.}
\label{fig:twoskeins}
\end{figure}
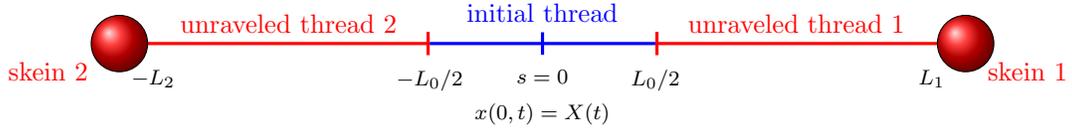

Another scenario of unraveling is when a skein splits into smaller connected fractions, which then unravel. Here we consider the simple case of a skein breaking into two halves. The unraveling may be faster since the initial viscous drag is dominated by two skeins, rather than a skein and a small initial length of thread. A diagram of this configuration is show in Fig.~\ref{fig:twoskeins}: we model the broken skein as two spheres, of radius~$\R_1$ and~$\R_2$ respectively, connected by an unraveled length of thread, which can unspool at both ends.  We fix a reference point~$\s=0$ between the two skeins such that~$\xc(0,\t)=\X(\t)$.  The thread then extends a length~$\L_1(\t)$ towards the first skein (right) and~$\L_2(\t)$ towards the second skein (left), with~$\L = \L_1 + \L_2$ the total unraveled length.  Without loss of generality we take~$\L_1(0) = \L_2(0) = \tfrac12\L_0$. The peeling force at the first skein ($\s=\L_1$, $\xc = \X+\L_1$) is the sum of the drag forces due to the second skein ($\s=-\L_2$, $\xc = \X-\L_2$) and drag on the thread:
\begin{equation}
  \T|_{\s=\L_1}
  =
  -6\pi\dvisc\R_2(\uc(\X-\L_2,\t) - (\dot\X - \dot\L_2))
  -4\pi\dvisc\L\delta\,(\bar\uc(\L,\X,\t) - \dot\X)
  \label{eq:Tskein1}
\end{equation}
where now $\bar\uc(\L,\X,\t) \ldef \frac{1}{\L_1+\L_2} \int_{\X-\L_2}^{\X+L_1}\uc(\xc,\t)\dint\xc$.  Since the thread and skeins are free, the peeling force at $\s=L_1$~\eqref{eq:Tskein1} must balance the viscous drag force on the first skein:
\begin{equation}
  \T|_{\s=\L_1}
  =
  6\pi\dvisc\R_1(\uc(\X+\L_1,\t) - (\dot\X + \dot\L_1)).
  \label{eq:Tskein1b}
\end{equation}
Equating~\eqref{eq:Tskein1} and~\eqref{eq:Tskein1b}, we can solve for~$\dot\X$:
\begin{equation}
  \dot\X
  =
  \frac{
  3\R_1(\uc(\X+\L_1,\t) - \dot\L_1)
  +
  3\R_2(\uc(\X-\L_2,\t) + \dot\L_2)
  +
  2\L\delta\,\bar\uc(\L,\X,\t)
  }
  {3(\R_1 + \R_2) + 2\L\delta}\,.
\end{equation}
We use this to eliminate~$\dot\X$ from~\eqref{eq:Tskein1b}:
\begin{multline}
  \T|_{\s=\L_1}
  =
  \frac{6\pi\dvisc\R_1}{3(\R_1 + \R_2) + 2\L\delta}
  (3\R_2(\uc(\X+\L_1,\t) - \uc(\X-\L_2,\t) - \dot\L)
  \\
  + 2\L\delta
  (
  \uc(\X+\L_1,\t)
  -
  \bar\uc(\L,\X,\t) - \dot\L_1)
  ).
  \label{eq:TL1}
\end{multline}
We can then also carry out the same calculation for the second skein, at~$\s=-\L_2$, and find
\begin{multline}
  \T|_{\s=-\L_2}
  =
  \frac{6\pi\dvisc\R_2}{3(\R_1 + \R_2) + 2\L\delta}
  (3\R_1(\uc(\X+\L_1,\t) - \uc(\X-\L_2,\t) - \dot\L)
  \\
  - 2\L\delta
  (
  \uc(\X+\L_2,\t)
  -
  \bar\uc(\L,\X,\t) + \dot\L_2)
  ).
  \label{eq:TL2}
\end{multline}
Now it is a matter of solving the coupled peeling equation~$\FVconst(\dot\L_1)^\FVpow = \T|_{\s=\L_1}$, $\FVconst(\dot\L_2)^\FVpow = \T|_{\s=-\L_2}$.  To keep things simple, let us take a symmetric configuration centered on~$\xc=\X=0$ where the two skeins are initially of equal size. (Unequal splitting would result in a depletion time in between this case of even splitting and the free skein-thread of Sec.~\ref{sec:freeskeinthread}.)  We take an antisymmetric velocity field~$\uc(\xc,\t) = -\uc(-\xc,\t)$ that pulls apart the skeins, such as for an extensional flow~$\uc=\lambda\xc$.  Then~$\R_1=\R_2$ and~$\L_1=\L_2$ for all time, and~$\bar\uc=0$. The tensions~\eqref{eq:TL1} and~\eqref{eq:TL2} are then equal and greatly simplify to $ \T|_{\s=\L_1} = \T|_{\s=-\L_2} = 6\pi\dvisc\R_1\,(\uc(\L_1,\t) - \dot\L_1)$. Thus, the dynamics for this case is governed by
\begin{equation}
  \FVconst(\dot\L_1)^\FVpow =
  6\pi\dvisc\R_1\,(\uc(\L_1,\t) - \dot\L_1).
  \label{eq:symmskeins}
\end{equation}
The drag force on the thread has dropped out, since the antisymmetric velocity field leads to canceling forces on the thread.  Another way to think of~\eqref{eq:symmskeins} is to observe that in making a symmetric configuration, with the two skeins being pulled apart by a straining flow centered on the origin, we have effectively `pinned' the thread at~$\xc=0$. We have thus recovered our pinned thread equation~\eqref{eq:pinnedthread} from Sec.~\ref{sec:pinnedthread}, with the notable difference that now we cannot use a constant velocity field~$\U$, but must resort to a straining flow~$\lambda\xc$ or some other nonuniform flow.

We nondimensionalize \eqref{eq:symmskeins} using a characteristic length scale $\R_0$ and  obtain \begin{equation}
  (\dot\L_1^*)^\FVpow =
  \wp \R_1^*\,(\uc^*(\L_1^*,\t^*) - \dot\L_1^*),
\end{equation}
where  $\dot\L^* = {\dot{\L}}/{\lambda \R_0}$ and $\R^* = {\R}/{\R_0}$ are the nondimensional unraveling rate and skein radius, respectively. The natural dimensionless number in this case is
\begin{equation}
  \wp = \frac{6\pi\dvisc\,\lambda\,\R_0^2}{\FVconst\,(\lambda\,\R_0)^{\FVpow}}
  = 6\pi\dvisc\,\R_0^{2-\FVpow}\,\lambda^{1-\FVpow}\,\FVconst^{-1}.
  \label{eq:Pcase4}
\end{equation}

\begin{figure}
\begin{center}
  \includegraphics[width=.7\columnwidth]{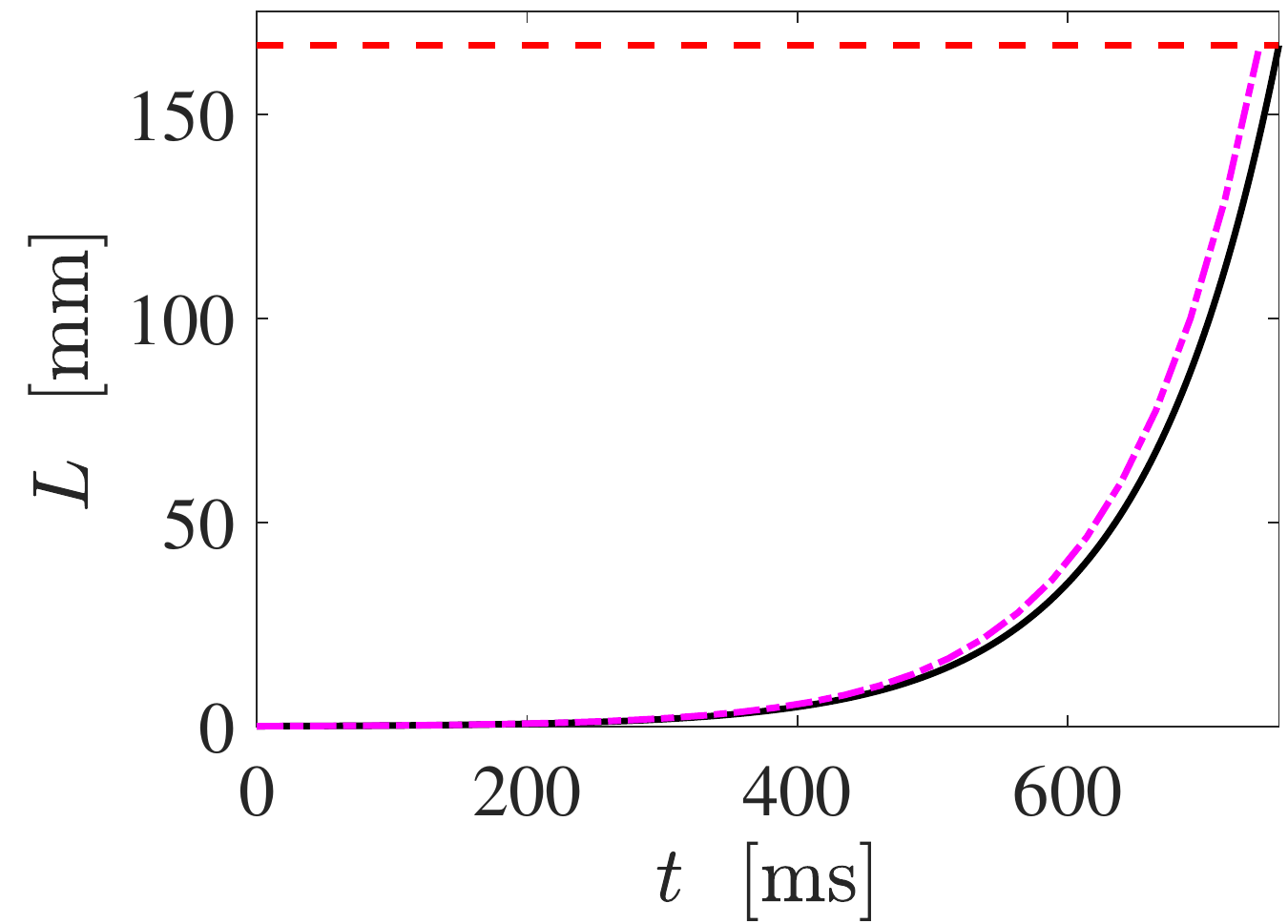}
\end{center}
\caption{Numerical solution (solid line) for the thread half-length~$\L_1(\t)$ using the force for two symmetric free skeins (Eq.~\eqref{eq:symmskeins}) for the parameter values $\R_1(0)=50 \,\microm$, $\L_1(0)=2\R_1(0)$, $\wp = 10$, $\FVpow=1/2$, $\lambda = 10\,\second^{-1}$.  The \textbf{-- ---} line is the upper bound~$\L_1(0)\exp(\lambda\t)$.  The horizontal dashed line is at~$\L_1={\L_1}_{\text{max}}$, when the skein is fully unraveled.}
\label{fig:Lsolve_two_skeins}
\end{figure}

Figure~\ref{fig:Lsolve_two_skeins} shows the unraveling dynamics associated with a split skein using parameter values similar to the free configuration of Sec.~\ref{sec:freeskeinthread} and Fig.~\ref{fig:Lsolve_freefree}. The free skein unravels  faster when split, as expected ($0.756\,\second$ vs $1.73\,\second$), owing to a stronger effective drag force and a kinematic upper bound with a rate~$\lambda$ rather than~$\tfrac12\lambda$. There is an important difference between using the free thread-skein equation~\eqref{eq:freefree} and the free split-skein equation~\eqref{eq:symmskeins}: the former has a drag slaved to a short initial thread length, whereas for the latter the drag depends on the initial radius of the split skein, which can easily be larger.

In addition to the four cases discussed in this section, we also analyzed a slightly more realistic scenario of suction flow where velocity decays away from the mouth of the predator and we consider a pinned skein at different locations away from the mouth. We use an approximate flow profile from the experimental data available in the literature (S.I.\ Sec.~\ref{sec:suction}). In general, such a flow profile is both spatially and temporally varying, but we neglect time-dependent variations for our analysis. The peak velocity (at the mouth of predator) was chosen to match the characteristic velocity ($\U=1\,\meter/\second$). The velocity decays over a characteristic length scale on the order of the gape size (e.g.~opening size of the mouth). We estimate gape size from the video evidence \citep{Zintzen2011} of slime deployment, resulting in extensional strain rates between $0.28$--$2.2\,\second^{-1}$), with the rate being highest at the predator mouth. These are smaller extension rates than considered in the earlier cases of this section. The choice of pinning location drastically affects the unraveling time.  A depletion time of $\sim0.4\,\second$ was obtained for the case where the skein is pinned at a distance equal to 1/3 of the gape size of the predator. This is longer than the unravel time we found for the pinned skein in a uniform flow ($\approx 0.22\,\second$), due to the decaying velocity away from the predator's mouth, but the unraveling time still falls close to natural unraveling time scales. More complicated spatially and temporally varying flow fields can be treated in a similar way; we expect the timescales in such cases to be on the order of those we found, given that in real scenarios the thread-skein system can be very close to or in the mouth of the predator, and the nondimensional quantity $\wp$ is likely to be sufficiently large.

\section{Discussion}
\label{sec:discussion}

\subsection{The role of the dimensionless parameters \texorpdfstring{$\wp$}{P} and \texorpdfstring{$\FVpow$}{m}}

The unraveling times for various cases discussed in Sec.~\ref{sec:skeinflow} depend on the dimensionless quantity $\wp$, but also separately on the model parameter~$\FVpow$ in the peeling force law. This is clear from the dimensionless governing equations in Sec.~\ref{sec:skeinflow} that depend on these two dimensionless parameters separately, although \FVpow~also appears in the definition of $\wp$. The power-law exponent~$\FVpow$ determines the peeling force dependence on the unraveling rate. Such a rate dependence exists in peeling scenarios due to the viscoelastic nature of adhesion at the peeling site. In the case of hagfish thread peeling from the skein, the dependence can possibly arise from viscoelastic timescales involved in the deformation of mucous vesicles or the polymeric solution of mucus \cite{Boni2016}, or the protein adhesive between the loops of thread~\cite{bernards2014}. The peeling resistance also depends on the dimensional constant factor, $\FVconst$, but its influence on unraveling is built into the dimensionless factor $\wp$, for which $\wp\sim \FVconst^{-1}$.

\begin{figure}[t]
    \centering
    \includegraphics[width=\textwidth]{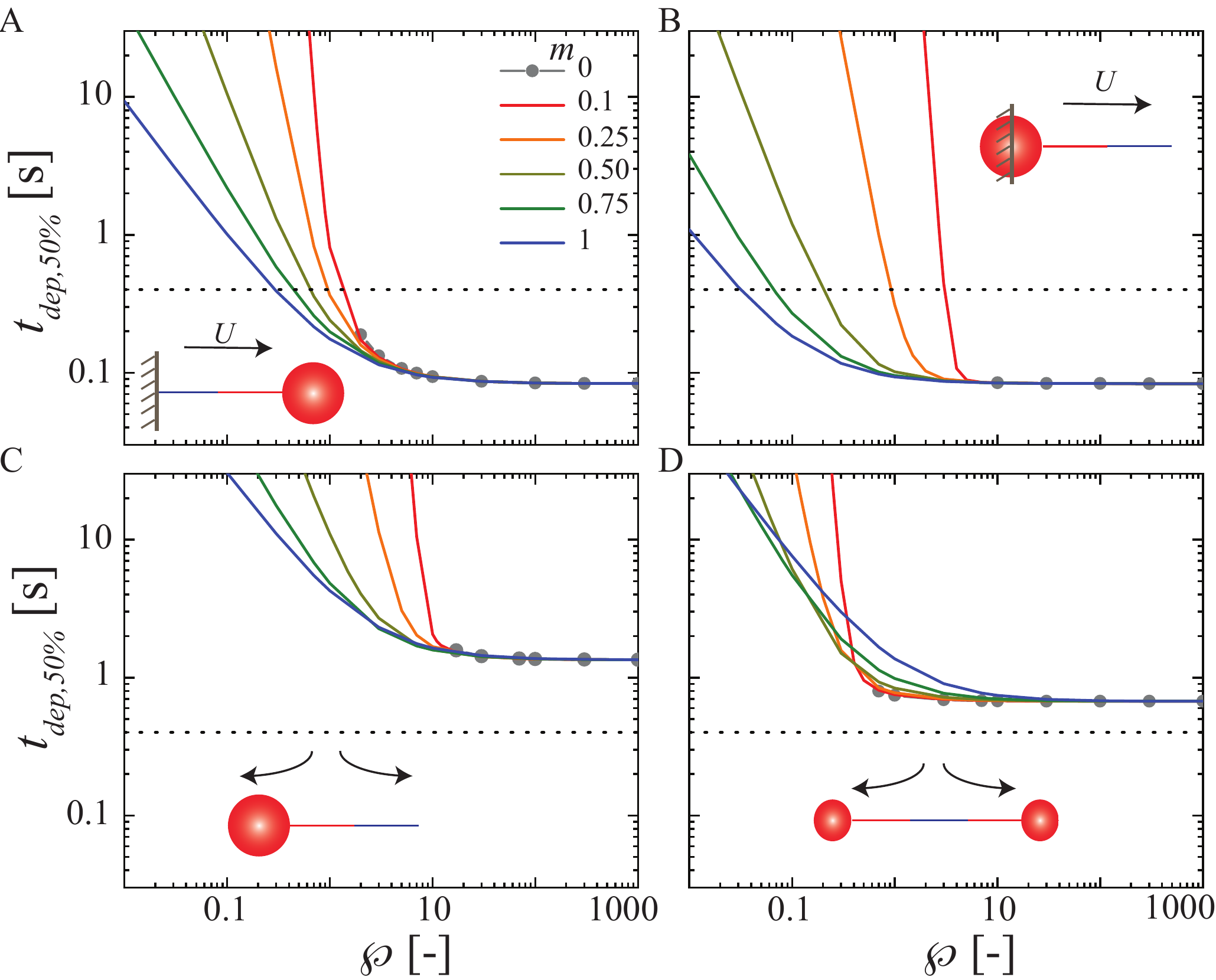}
    \caption{Parameter dependence of ``effective" unraveling. Comparison of timescale~$\tdepfifty$ for unraveling half the total length of the fiber for different values of $\FVpow$, as the dimensionless quantity~$\wp$ is varied in different unraveling scenarios. (A) pinned thread in uniform flow, (B) pinned skein in uniform flow, (C) free thread and skein in straining flow, and (D) symmetric free skeins in straining flow. Other parameters used are $\rr =1 \,\microm $, $\R_0=50 \,\microm$, $\L_0=2\R_0$, $\U=1\,\meter/\second$, and $\lambda=10\,\second^{-1}$. The dotted horizontal line represents the physiologically observed timescale ($= 0.4\,\second$).}
    \label{fig:depletiontime_all}
\end{figure}

Figure~\ref{fig:depletiontime_all} compares all four flow scenarios of Sec.~\ref{sec:skeinflow} as a function of $\wp$ and $\FVpow$ in terms of the effective deployment time, i.e.~the time to unravel half of the thread length, $\tdepfifty$. (This effective time is used because some flows cannot fully deplete the skein in finite time for $\FVpow<1/3$, as discussed in Sec.~\ref{sec:skeinflow}.  Moreover, in practice the threads do not need to be fully unraveled to create slime.) The flow parameters are identical to those previously described. For all cases, the limit of high drag and low peel resistance, $\wp\gg 1$, converges to the kinematic limit of unraveling where unconstrained portions of the skein-thread system exactly advect with the local flow velocity. At the other extreme, viscous drag is weak compared to peel resistance and for some small value of $\wp$ unraveling is too slow to match physiological timescales.

The power-law exponent $\FVpow$ is a secondary effect compared to $\wp$. In general, for $\wp>10$, $\FVpow$ has negligible effect on unraveling times.  For $\wp<10$, the dependence on $\FVpow$ is case-specific. For the cases of pinned thread  (Sec.~\ref{sec:pinnedthread}), pinned skein (Sec.~\ref{sec:pinnedskein}), and free skein-thread (Sec.~\ref{sec:freeskeinthread}), a larger value of $\FVpow$ leads to a smaller unraveling time (while keeping the same value of $\wp$). For the case of skein splitting in an extensional flow (Sec.~\ref{sec:twoskeins}), such a monotonic trend is not observed and above a critical value of $\wp$ the unraveling is faster for small values of $\FVpow$. This presumably arises from the nonlinearity in the peel force constitutive equation. For example, taking $\FVpow=1$ as a reference case, making $\FVpow<1$ \emph{increases} the dimensionless peel resistance for $\dot{\L}^*<1$, but \emph{decreases} the peel resistance for $\dot{\L}^*>1$. As such, whether the unraveling rate is $\dot{\L}^*\gtrless 1$, the exponent $\FVpow$ can accelerate or decelerate the unraveling process.

The value of $\FVpow$ affects the minimum required $\wpmin$ to achieve unravel times comparable to physiological timescales (i.e.~$\tdepfifty$ at or below the dotted lines in Fig.~\ref{fig:depletiontime_all}). For the uniform velocity field cases ($\U=1\,\meter/\second$), $\wpmin$ is a weaker function of $\FVpow$ for the pinned thread case, $\wpmin=0.29$--$1.32$, compared to the pinned skein case, $\wpmin=0.03$--$3$. For the cases of a free skein-thread in extensional flow, even with splitting, $\tdepfifty$ never falls below $400\,\millisecond$ even at high $\wp$. That is, $\tdepfifty$ is higher than the physiological unravel timescales by a factor of 2 or 3. However, as stated earlier, the timescales in such cases are determined by the specific choice of strain rate,~$\lambda$, and the initial unravel length~$\L_0$. Such kinematic and geometric parameters are certainly variable in reality, and small changes could easily decrease the unravel timescales, as previously discussed. In any case, $\FVpow$ becomes important only when depletion timescales are much larger than the kinematic limit, clearly showing that $\FVpow$ is of secondary concern compared to $\wp$.

An important caveat to the $\FVpow=0$ case of $\Fp=\FVconst=$~constant is that peeling cannot occur ($\dot{\L}=0$) if the viscous drag falls below a critical value. For example, if either the initial skein radius~$\R_0$ or thread length~$\L_0$ is too small, the viscous drag force is less than $\Fp$ and unraveling cannot occur. Thus, in Fig.~\ref{fig:depletiontime_all} a minimum value of $\wp$ is needed for the $\FVpow=0$ cases. The minimum values ranges from about 1--10, depending on the case and the corresponding definition of $\wp$ for the flow and geometry. In three cases, the viscous drag can potentially increase during unraveling as the thread elongates (Fig.~\ref{fig:depletiontime_all}B--D). In these cases, the minimum $\wp$ is associated with initiating the peeling process. For the other case of the pinned thread, Fig.~\ref{fig:depletiontime_all}A, the viscous drag decreases during unraveling, since it is slaved to the skein radius which decreases in size during the process. Unraveling here will eventually stop at a critical value of $\R$. This therefore feeds back to requiring a larger critical initial value of $\wp$ to unravel by 50\%, and is used in Fig.~\ref{fig:depletiontime_all}A to determine the domain of $\wp$ for the $\FVpow=0$ case.

\subsection{Estimating the parameter \texorpdfstring{$\wp$}{P}}

A key question remains: what is $\wp$ in physiological scenarios? For this we must know the peeling force parameters in the constitutive model, and no direct experimental measurements are yet available. Here we make estimates for the two extreme conditions of $\FVpow=1$ and $\FVpow=0$, i.e.~a linear dependence on velocity (akin to a constant viscous damping coefficient)  and a constant peel force, respectively.

For the $\FVpow=1$ case, we consider viscous resistance acting at the peel site with stress $\sigma=\dvisc_u\dot\epsilon$, where $\dvisc_u$ is the uniaxial extensional viscosity between the separating thread and the skein (related to shear viscosity as $\dvisc_u=3 \dvisc$), and $\dot\epsilon = \dot{\L}/\Lc$ is the local extensional strain rate that depends on the peel velocity $\dot{\L}$ and the characteristic velocity gradient length $\Lc$. The stress acts over the characteristic thread-thread contact area, which we assume scales as $A\approx \dd^2$, i.e.~contact across the diameter and the length of contact along the thread also scales with the diameter. The peel force is then
\begin{equation}
\Fp \approx \dvisc_u \bigl( \nofrac{\dot{\L}}{\Lc} \bigr)\dd^2.
\end{equation}
Comparing to the peeling law in \eqref{eq:FpV}, $\Fp =\FVconst \dot{\L}^\FVpow$, we get
\begin{equation}
  \FVconst = \dvisc_u \nofrac{\dd^2}{\Lc}; \qquad \FVpow=1 .
\end{equation}
Substituting this into $\wp=\Fd / \Fp$, and considering the majority of cases where drag is set by the skein radius $\R_0$, i.e.~equations~\eqref{eq:Pcase1}, \eqref{eq:Pcase3}, \eqref{eq:Pcase4}, we obtain
\begin{equation} \label{eq:Pest_m1}
    \wp =  \frac{6\pi\dvisc\R_0\,\U}{\dvisc_u \bigl( \nofrac{\dot{\L}}{\Lc} \bigr)\dd^2}
    = 6 \pi \frac{\dvisc}{\dvisc_u} \frac{U}{\dot{\L}} \frac{\R_0\Lc}{\dd^2}
\end{equation}
where important ratios have been grouped. The simplest case is peel viscosity arising from the surrounding viscous liquid at the peel site. In other words, the viscosity causing drag is also resisting peeling, and to cast in terms of extensional viscosity we take $\dvisc_u=3 \dvisc$, a result for a Newtonian fluid. The viscosity $\dvisc$ may be that of sea water, or a surrounding mucous vesicle solution with higher viscosity, but under these assumptions the ratio $\dvisc / \dvisc_u$ is still the same. Furthermore, typically $U/\dot{\L}\gtrsim 1$, and the velocity gradient length scale is likely set by the thread radius, $\Lc\approx \rr$. Then~\eqref{eq:Pest_m1} dramatically simplifies to
\begin{equation}
    \wp = 2 \pi \nofrac{\R_0}{\dd}
\end{equation}
which clearly estimates $\wp \gg 1$, or more specifically for $\R_0=50\,\microm$ and $\dd=2\,\microm$, $\wp \approx 160$. If drag is instead dominated by the thread length $\L_0$, e.g.~for the pinned skein case of Sec.~\ref{sec:pinnedskein}, then the numerator in~\eqref{eq:Pest_m1} would be modified by replacing $6 \pi \R_0$ with $4 \pi L_0$. We expect $\L_0$ to be the same order as $\R_0$, e.g.~a single curl in the coil. But if $\L_0$ is smaller, it would decrease $\wp$ accordingly.

\begin{figure}
    \centering
    \includegraphics[width=\textwidth]{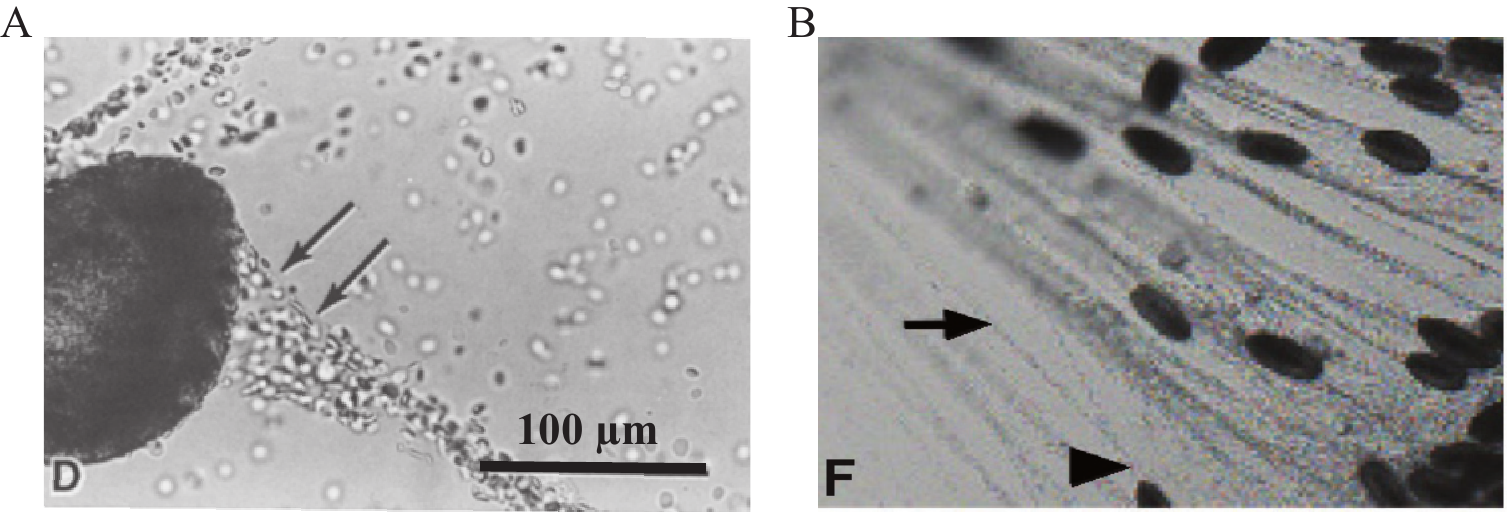}
    \caption{(A) Mucous vesicles aggregating on unraveling thread (adapted from Koch et al.~\cite{koch1991}). (B) Mucous vesicles aggregated on unraveling thread elongated along with the fiber under the flow (adapted from Winegard et al.~\citep{Winegard2010}).}
    \label{fig:kochwinegard}
\end{figure}

Our specific assumptions can modify the details, but in general we estimate that physiological conditions for $\FVpow=1$ would give $\wp >1$, if not $\wp \gg 1$. The velocity gradient length scale $\Lc$ could be smaller than the thread radius $\rr$. A decrease in $\Lc/\rr$ makes $\wp$ proportionally smaller, but it is difficult to imagine this being more dramatic than, say, a factor of 10. The velocity ratio $U/\dot{\L}$, if anything, will be larger than 1, and this proportionally increases the estimate of $\wp$. The viscosity ratio $\dvisc / \dvisc_u$ could be smaller, e.g.~if the viscosity of proteins between thread wrappings is larger than the surrounding viscous liquid. However, we note that the surrounding viscous liquid can have a very large viscosity, e.g.~the measured extensional viscosity of hagfish mucous vesicle solutions obtained by B\"{o}ni et al.~\citep{Boni2016} is $\dvisc_u \approx 10\,\Pa\cdot\second$. This is much higher than water, $\dvisc_u\approx3\,\milliPa\cdot \second$. An additional mechanism of increasing drag, and $\wp$, is for mucous vesicles to bind on the thread during the unraveling process (Fig.~\ref{fig:kochwinegard}), which would transmit additional forces to the drag term, as suggested by Winegard et al.~\citep{Winegard2010}. Such a scenario is possible since mucous vesicles and threads cells are densely packed inside the slime glands and are released simultaneously. For all of these variations, $\wp >1$ seems very likely for physiological conditions in this constant viscosity estimate for $\FVpow=1$.

To estimate physiological $\wp$ for $\FVpow=0$, the other extreme of a constant force resisting peel, we consider peel strength interactions between the skein fibers  solely due to van der Waals forces. We estimate $\Fp = \FVconst \approx 10^{-7}\,\Newton$  (see S.I.\ Sec.~\ref{sec:force}). Substituting into $\wp=\Fd / \Fp$, and considering cases where drag is set by the skein radius, i.e. Eqs.~\eqref{eq:Pcase1}, \eqref{eq:Pcase3}, \eqref{eq:Pcase4}, with $\R_0 = 50~\microm$, water viscosity $\dvisc=1\,\milliPa\cdot\second$, and $U=1\,\meter/\second$, gives $\wp\approx 90$. We see that~$\wp \gg 1$ with these assumptions. Even if the force resisting peeling is larger by a factor of 10 or 100, still $\wp \gtrsim 1$ and viscous hydrodynamics can provide rapid unraveling that can be very close to the kinematically-derived lower bounds on unraveling time.

\section{Conclusion}

Our analysis shows that, under reasonable physiological conditions, unraveling due to viscous drag can occur within a few hundred milliseconds and is accelerated if the skein is pinned at a surface, such as the mouth of a predator. A dimensionless ratio of viscous drag to peeling resistance, $\wp = \Fd / \Fp$, appears in the dynamical equations and is the primary factor determining unraveling timescales. Large $\wp$ corresponds to fast unraveling that approaches a kinematic limit wherein free portions of the thread-skein system directly advect with the local flow velocity. For characteristic velocity $\U$, the bound is $\tdep \ge \Lmax/\U$, whereas for extensional flows with strain rate $\lambda$, $\tdep \ge \lambda^{-1}\log(\Lmax/\L_0)$, where~$\L_0$ is the initial thread length.

The modeling approach captures essential features and insight by considering a single dilute skein unraveling in idealized flow fields.  Future modeling efforts could build on our work by expanding and detailing several aspects, primarily with new experimental measurements of peel strength for skein unraveling, but also details of physiological flow fields including characteristic velocities and strain-rates. Real physiological scenarios are more complex due to chaotic flows and multi-body interactions (multiple skeins, mucous vesicles). Our model does not consider such interactions, or the important feature of unraveled threads interacting to create a network. At leading order, we expect such modeling to require more complex flow fields that create extension (to unravel fibers) but also bring different fibers together. Mixing flows would be excellent candidates for theoretical analysis, and any experimental characterization of physiological flow fields should keep this perspective in mind, e.g.~simple suction flow with extension, but no mixing, may not be sufficient to create a network of unraveled threads.

Although the physiological flow fields may be different from the ones that were used in the analysis, our results underline the importance of viscous hydrodynamics and boundary conditions on the process. Recent work \citep{bernards2014} found that Pacific hagfish thread cells undergo spontaneous unraveling in salt solution. However, the unraveling timescales ($\sim$~min) are much larger than the in physiological timescales ($\sim 0.4\,\second$) during the attack. It is possible that ion transport to the peeling site may help in peeling the adhesive contacts, which may be diffusion-limited without flow. Although it is known that flow is required to accelerate unraveling to $\tdepfifty <1\,\second$, it is not yet clear whether flow-enhanced ion transport may also contribute to a faster unraveling, in addition to the drag effects. If so, this would modify the $\Fp$ behavior and require modeling of transport at the peel site due to flow. Our results do not rule out the possibility of ion-mediated unraveling but provides an alternate mechanism of unraveling which may be occurring alone or in conjunction with a multitude of other processes.


\section*{Acknowledgments}
We thank Prof.\ Douglas S.\ Fudge (Chapman University, CA, USA) for providing the hagfish exudate and for many helpful discussions. GC and RHE are grateful to the National Science Foundation for supporting their contributions to this research under grant number CBET-1342408.

\medskip
\centerline{\bf\small REFERENCES}
\bibliographystyle{abbrv}
\bibliography{hagfish}



\pagebreak


\begin{center}
  \textbf{\large Supplementary Information: Unraveling hagfish slime}
\end{center}


\setcounter{equation}{0}
\setcounter{figure}{0}
\setcounter{section}{0}
\setcounter{table}{0}
\setcounter{page}{1}
\renewcommand{\theequation}{S\arabic{equation}}
\renewcommand{\thefigure}{S\arabic{figure}}
\renewcommand{\bibnumfmt}[1]{[S#1]}

\section{Materials and microscopy}
\label{sec:materials}

Pacific (\textit{Eptatretus stoutii}) and Atlantic (\textit{Myxine glutinosa}) hagfish were maintained at the University of Guelph as previously described \citep{Winegard2010}. All housing, feeding, and experimental conditions were approved by the University of Guelph Animal Care Committee (Animal Utilization Protocol 2519). Hagfish were anesthetized
before the extraction of exudate. Electrical stimulation near the slime glands resulted in the secretion of the exudate locally near the site of stimulation \citep{Winegard2010}. The exudate was collected directly from fish using a spatula, stored in a microcentrifuge tube at $10^{\circ}\mathrm{C}$. No buffer or oil was used to stabilize the exudate from Pacific hagfish and it was used within 2--5 hours from the time of extraction to ensure sample viability. The exudate extracted from Atlantic hagfish was stabilized under mineral oil and stored on ice for transport to the testing location \citep{Herr2010}.

For microscopy studies, an inverted optical microscope with brightfield imaging was used.  To visualize a single skein unraveling, a very small volume of Atlantic hagfish exudate was added to a pool of artificial sea water and an isolated skein was located and observed under a flow created by manually disturbing the sample. For the network images, a precise volume of exudate from Pacific hagfish was introduced into a centrifuge tube containing artificial sea water ($1\,\mathrm{ml}$) and the contents were sloshed back and forth in the closed tube six times, similar to prior work~\citep{Ewoldt2011}. Following this the sample looked like a cohesive mass. Gel samples of various exudate concentrations were transferred to glass slides with small wells. A Nikon 90i Eclipse microscope was used to visualize the network with differential interference contrast (DIC) to image the slime network.

\section{Minimum peeling force}
\label{sec:force}
\begin{figure}
\centering
\begin{tikzpicture}

\draw[black, very thick] (3.2,0) -- (7,0);
\draw[red, very thick] (6,-0.02) -- (7,-0.02);

  \draw[black, thick,<->] (6,0.15) -- (7,0.15) node[pos=0.5,sloped,above]{$\delta $};
\draw[red, thick] (6,-0.02) parabola (3,-2);

\draw[black] (4.8,0) arc (180:220:0.5) node[pos=0.43,rotate=90,sloped,left]{$\theta$};

\shadedraw[ball color=red] (3,-2) circle (.05);
\draw[black, thick,->] (2.8,-2.4) -- (2.3,-3.2) node[pos=.5,sloped,below]{$\Fp$};
\draw[black, thick,->] (2.6,-2) -- (2.3,-2.4)
node[pos=.5,sloped,above]{$\deltap$};

\end{tikzpicture}
\caption{Schematic of a fiber (red) peeling from a substrate (black).}
\label{fig:peeling}
\end{figure}
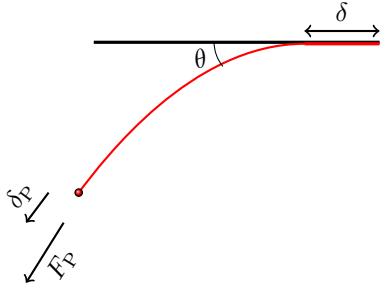

Assuming only van der Waals interactions exist at the peeling site, we estimate of the minimum peeling force required for peeling. For an inextensible fiber, with negligible bending energy storage during peeling, the work done by the peeling force~$\Fp$ in moving a point on the fiber by a distance~$\deltap$ is~$\Fp\, \deltap$ (schematic in Fig.~\ref{fig:peeling}). The energy involved in creating new surface at the peeling site is equal to the van der Waals interaction energy released at the peeling site, given by $E_A \Ap$, where $E_A$ is adhesion energy per unit area for van der Waals surfaces. The area created at the peeling site is $\Ap\approx (2\rr)\,\delta$, where~$\rr$ is the thread radius and~$\delta$ is the displacement length of the peeling front created by the fiber displacement~$\deltap$. The energy balance, rearranged, gives $\Fp = 2E_A\,\rr\, (\nofrac{\delta}{\deltap})$. For a wide range of peeling angles $\theta$, $\deltap\sim \delta$ (note that $\deltap =\delta$ for $\theta=\pi/2$), and we get $\Fp \sim 2 E_A r$. A typical adhesion energy for van der Waals surfaces is $50$--$60\,\mJoule/\meter^2$ \citep{Autumn2002,Kirmse2010} and $\rr\approx 1\,\microm$, which gives $\Fp \approx 0.1\,\microNewton$.

\section{Hagfish defense in suction flow}
\label{sec:suction}

\begin{figure}
\begin{tikzpicture}

\node at (0,11.5) {(A)};
\node[inner sep=0pt] (suction) at (8,9.5)
    {\includegraphics[width=\textwidth]{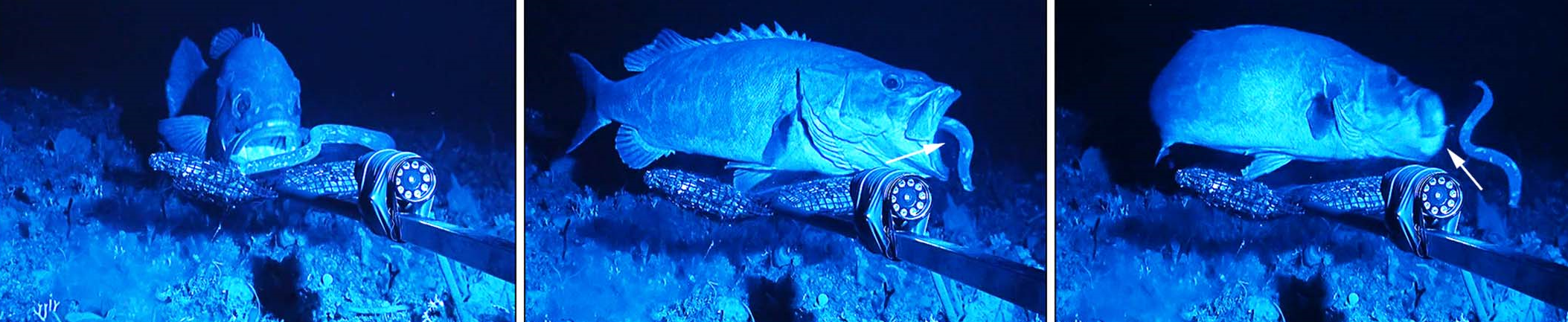}};

\node at (0,7.25) {(B)};

\node[inner sep=0pt] (suction) at (3.7,4.6)
    {\includegraphics[width=0.45\textwidth]{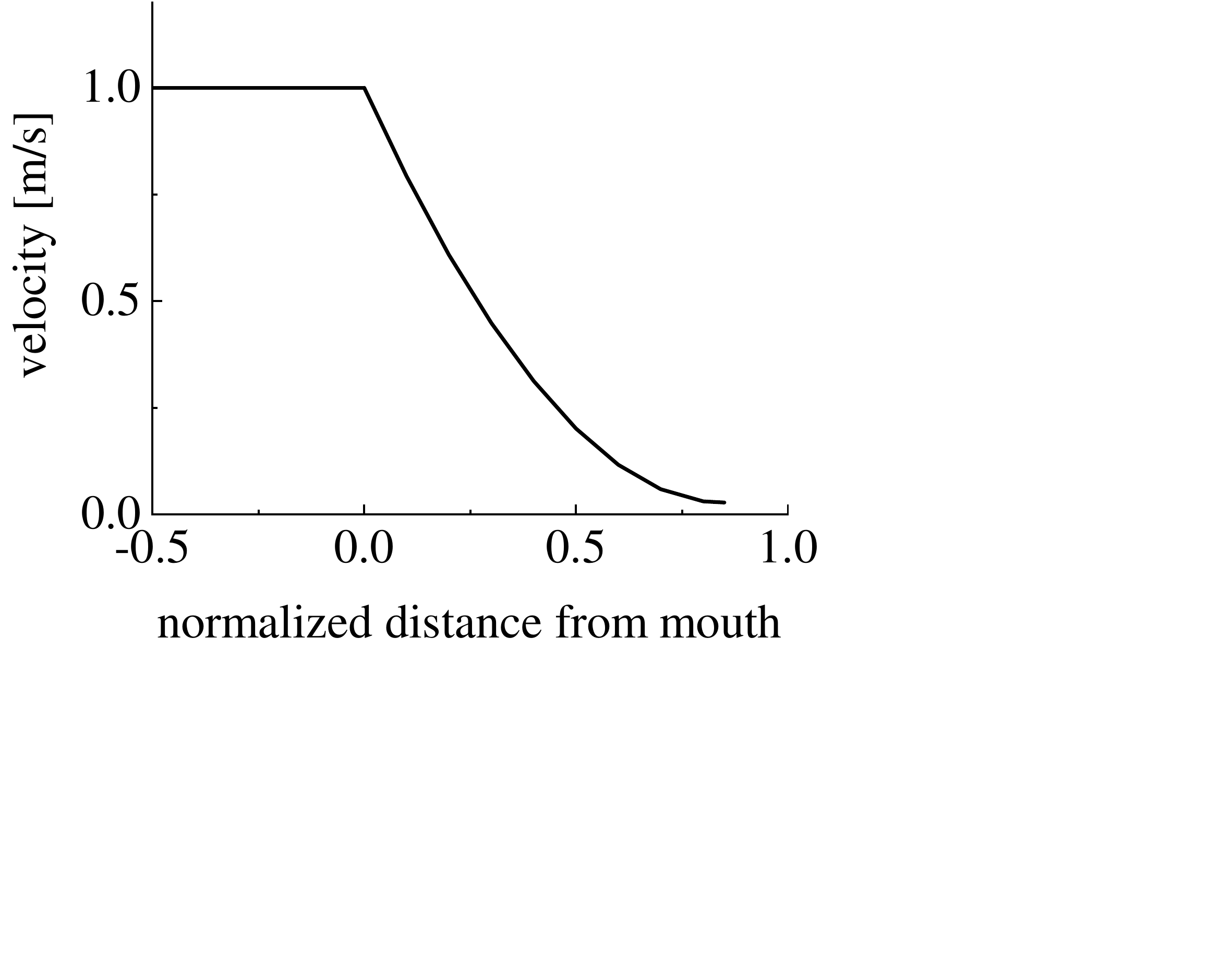}};

\begin{scope}[shift={(17,3.35)},xscale=-.8,yscale=.8]
  \begin{scope}[shift={(2.5,.78)},scale=.6]
    \draw[fill=cyan] (8,2) to[bend left=50] (13,2) to[bend left=50] (8,2);
    \draw[fill=cyan] (13,2) -- (14.5,3) -- (13.9,2) -- (14.5,1) -- cycle;
    \draw[fill=white] (8.6,2.1) circle (.2cm);
    \draw[fill=blue] (8.55,2.1) circle (.1cm);
  \end{scope}
  \draw[very thick,-,red] (4,2) -- (7,2);
  \shadedraw[ball color=red] (3.75,2) circle (.25);
\end{scope}

\end{tikzpicture}
\caption{(A) Evidence of hagfish (\textit{Eptatretus sp.}~2) defense against suction feeding predator (\textit{Polyprion americanus}, Atlantic wreckfish) (image adapted from \citep{Zintzen2011}). A suction feeding fish induces flow into its mouth by expanding its buccal cavity and drawnig its prey along with the flow. We hypothesize the suction flow to aid in unraveling of thread cells and set up the slime network. (B) The one-dimensional suction flow profile~\eqref{eq:suctionprofile} with an assumption of constant velocity inside the predator's mouth. (C) A schematic of the fish-thread-skein system (not to scale).}
\label{fig:suction_schematic}
\end{figure}

Evidence of hagfish defense with slime secretion is available for both biting and suction feeding predators \citep{Zintzen2011, Boni2016}. Here, we treat the case where the predator creates a suction flow to engulf the prey. A good approximation of suction flow is provided by numerous experimental and computational studies \citep{Higham2006, Day2007, Van2009, Skorczewski2009, Day2015}.  A one dimensionless parametrization of an observed suction flow with bluegill fish is
\begin{equation}
    u(x^*_g,t) = u(x^*_g=0,t)(1 - 2.19 x^*_g + 1.86 x^{*2}_g - 0.70 x^{*3}_g + 0.09 x^{*4}_g)
\label{eq:suctionprofile}
\end{equation}
where $u(x^*_g=0,t)$ is the flow speed at the mouth of the predator and $x^*_g$ is the non-dimensional distance from the predator mouth normalized by the gape size (the diameter of the open mouth). In \eqref{eq:suctionprofile}, both the velocity \emph{and} the extensional strain rate decay for $x^*>0$ (away from the mouth of the predator).  The form~\eqref{eq:suctionprofile} represents a time-varying velocity profile at the mouth of predator. Note that \eqref{eq:suctionprofile} applies outside the mouth.  We do not have data for the flow profile inside the mouth, but for our purpose here we assume a constant centerline velocity equal to the peak velocity (velocity at the gape), as if the flow was sucked into a constant diameter tube with a negligibly small boundary layer (Fig.\ref{fig:suction_schematic}B). The velocity field is continuous, though the spatial derivative is not. This discontinuous strain rate is not a problem since we eventually integrate the velocity to get an average over the unraveled length. To simplify the analysis, we use a time-invariant version of this velocity profile. We note that \eqref{eq:suctionprofile} is obtained by polynomial fit to PIV data, and hence should only be used in the range of $x^*_g$ in which it makes physical sense, i.e.~monotonically decreasing $\nofrac{\partial u(x^*_g,t)}{\partial x^*_g}$, which is true only for $x^*_g < 1.36$. We use a gape length of $10\,\cm$, an approximate scale obtained from Fig.\ref{fig:suction_schematic}A.

We solve for the case of a skein pinned at different locations~$x^*$ where the unraveling force is due to the drag on the unraveled fiber under the suction flow given by Eq.~\eqref{eq:suctionprofile} (Fig. \ref{fig:suction_schematic}C). The governing equations are the same as derived in the paper:
\begin{equation}
  (\dot\L)^\FVpow 
  = -4\pi\dvisc\,\FVconst^{-1}\L\,\delta(\L)\,
  \bigl(\dot\L + \bar\uc(\L,\X,\t)\bigr).
  \label{eq:pinnedskeinSI}
\end{equation}

\begin{figure}[ht]
    \centering
    \includegraphics[width=0.75\textwidth]{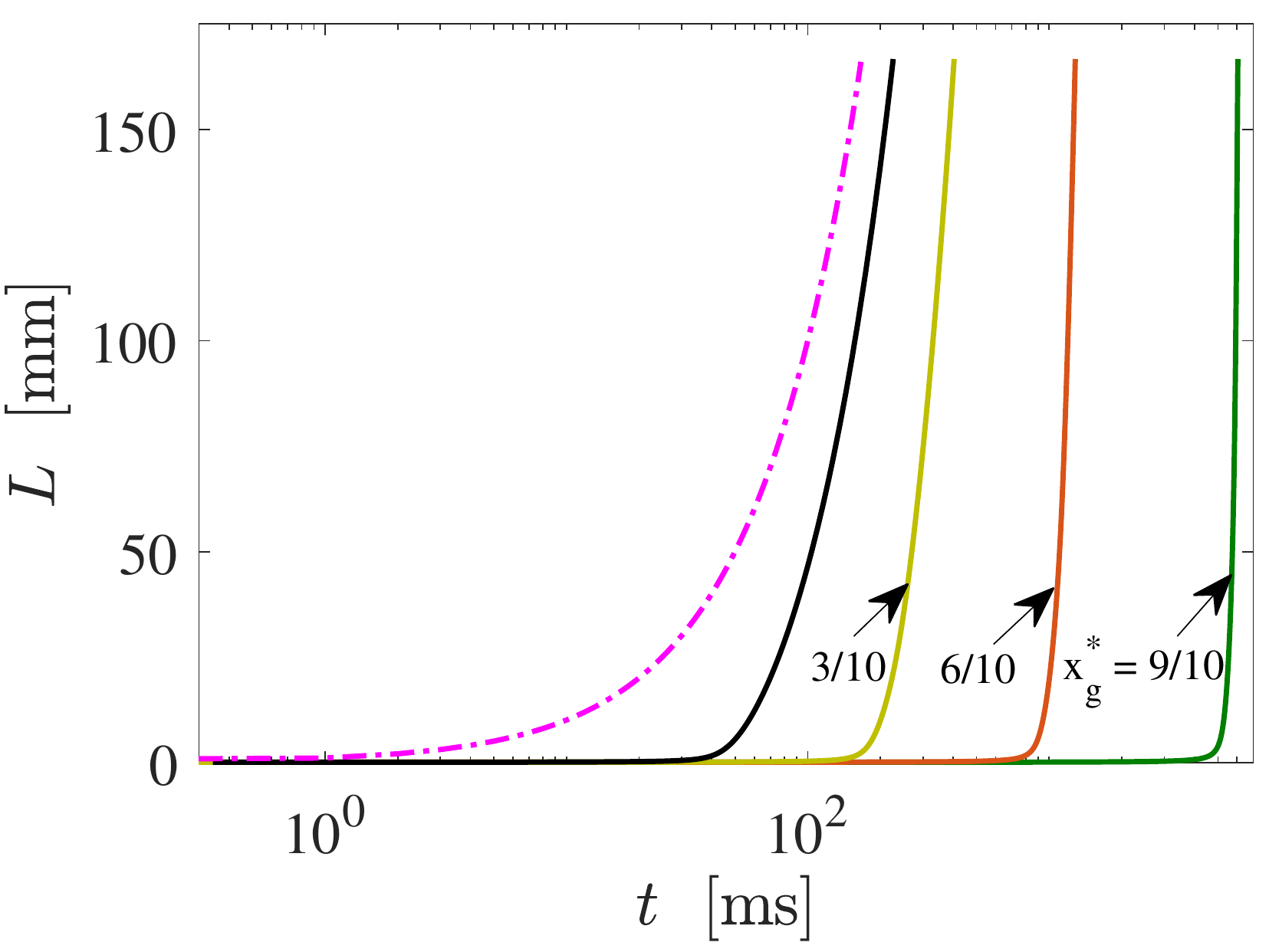}
    \caption{Numerical solution (dashed black line) of~\eqref{eq:pinnedskeinSI} for the parameter values $\R_0=50 \,\microm$, $\L_0=2\R_0$, $\wp = 0.5$, $\FVpow=1/2$, and velocity profile given by \eqref{eq:suctionprofile} with $u(x_g=0,t)=1\,\meter/\second$ with varying the location of the pinned point. Yellow, orange and green are the solution to the cases where the skin in pinned at locations $3/10,\,6/10$ and $9/10$ of the gape size ($=10\,\cm$). Solid black line is the numerical solution for the case of constant velocity of $1\,\meter/\second$. The dotted line is the upper bound~$\L = \L_0 + \U\t$, with $\U=1\,\meter/\second$.}
    \label{fig:suction_unraveling}
\end{figure}

Figure~\ref{fig:suction_unraveling} shows the numerical solution for this case. Compared to the case of uniform velocity field of $1\,\meter/\second$ solved in the paper, the unraveling time is longer with a spatially-varying velocity that decays away from the predator's mouth. The location of the pinning point of the skein has a huge impact on the unraveling timescales as shown in Fig.~\ref{fig:suction_unraveling}, since being farther from the predator means a lower flow velocity and extensional strain rate. The strain rate for the chosen velocity profile outside the predator's mouth lies between~$2.19\,\second^{-1}$ at~$x^*_g=0$ and~$0.27\,\second^{-1}$ at~$x^*_g=0.9$. From the video evidence of hagfish-predator attack, we note that exudate is released at a distance less than one-third of the gape size. This proximity to the mouth, in this flow field, results in an unraveling timescale close to the physiological one. Note that we used a velocity profile that was experimentally obtained for suction feeding fish with a gape size of only~$1.5\,\cm$ (bluegill). We therefore expect the magnitude of velocity in the real scenarios to be higher, and hence the unraveling times will be smaller for hagfish predators. It does pose the interesting possibility that smaller fish, with weaker suction flow, may not create a flow field that unravels the hagfish threads within hundreds of milliseconds.

\end{document}